# Constructing Linear Operators Using Classical Perturbation Theory [*]

Miguel Avillez [†] and David Arnas[‡]
*Purdue University, West Lafayette, IN 47907, USA*

**This work introduces a methodology for generating linear operators that approximately represent nonlinear systems of perturbed ordinary differential equations. This is done through the application of classical perturbation theory via the Lindstedt-Poincaré expansion, followed by an extension of the space of configuration that guarantees the linear representation of the expanded system of differential equations. To ensure that such a linear representation exists, this paper uses polynomial basis functions. Pseudo-code describing the implementation of the proposed method is listed. The method is applied to the Duffing oscillator as well as to the $J_2$ problem, with and without atmospheric drag, both analyzed using an osculating formulation. Additionally, conditions on the osculating Keplerian elements that produce low-eccentricity frozen orbits are presented, and a modification of the Lindstedt-Poincaré method is proposed to enable the generation of linear operators that dynamically adapt to changes in the frequency of the motion. Finally, the proposed method is compared with alternatives in the literature.**

## I. Introduction

Many problems in astrodynamics are related to the analysis or solution of initial value problems represented by nonlinear systems of ordinary differential equations. Examples of this include the motion of a satellite under the zonal-harmonics peturbation, the circular restricted three-body problem, and the relative motion between satellites. Since the analysis of nonlinear systems is generally very complex, they are often expressed as simpler linear systems, even if that representation is only valid in a small region of space. This allows taking advantage of the myriad of already-existing techniques for the analysis of linear systems, for example, to study their stability behavior, control possibilities, and estimation.

There are several methods for obtaining the linear representation of a nonlinear system. The most popular one is, of course, the first-order Taylor series expansion. However, this approach is only accurate in the close vicinity of the linearization point and over short propagation times. The next step in terms of complexity is the Carleman linearization, which transforms a system of differential equations into an infinite-dimensional linear system, by defining all monomials

---

of the original variables as new basis functions, and neglecting all terms that cannot be represented linearly [1]. Another method, the Koopman operator, is derived from operator theory and aims to transform a finite-dimensional nonlinear system into an infinite-dimensional linear system, which, for practical uses, is then truncated into a finite-dimensional space. Following the work of Mezić [2], the Koopman operator has been applied to a variety of fields, including fluid dynamics [3], control [4], and estimation [5]. All of these works were based on data-driven approximations of the Koopman operator. Although this type of procedure has been attempted in astrodynamics, its accuracy generally proved to be unsatisfactory [6], with the analytical evaluation of the Koopman matrix being necessary instead. This approach has been applied to attitude dynamics and control [7], to the zonal harmonics problem [8, 9], to orbits around a Lagrange point in the circular restricted three-body problem [6], to uncertainty propagation [10], and to the rendezvous problem [11].

In this work, we propose a new methodology based on classical perturbation theory, specifically the Lindstedt-Poincaré method, to create a linear operator that approximately represents a given nonlinear system. This is done by expanding the space of configuration, such that the differential equations resulting from the application of the Lindstedt-Poincaré method (which requires a power expansion) can be represented in linear form without any additional approximation. This provides the typical advantages associated with operator theory, that is, being able to apply techniques designed for the analysis of linear systems, and, compared to previous Koopman operators, allows representing the system linearly using much smaller matrices generated in significantly lower computation times. Additionally, the proposed method provides the advantages associated with classical perturbation theory, including the long-term stability of the approximated solution and the clear physical meaning of the nonlinear terms being neglected.

To showcase the performance and potential difficulties of the proposed method, we apply it to a very simple problem, the Duffing oscillator, and to two more complex systems in astrodynamics, the motion of a particle subject to the $J_2$ perturbation (a conservative problem), and the dynamics of an object subject to both the $J_2$ and atmospheric drag perturbations (a non-conservative problem). Since the $J_2$ problem has no general closed analytical solution, multiple approaches have been developed to generate approximate analytical solutions. Generally, these are obtained using some type of averaging, either of the Hamiltonian or of the used differential equations. In particular, Brouwer [12] and Garfinkel [13] followed the former approach (both using the von Zeipel perturbation method), while Kozai [14] followed the latter. Brouwer's solution was later improved by Kozai [15], Lydanne [16], Cohen and Lydanne [17], and more recently Lara [18], increasing the accuracy of the solution, extending the region where it is valid, and simplifying the methodology. A different perturbation method, based on Lie series and Lie transforms, was introduced by Hori [19] and Deprit [20, 21]. This type of method, used in numerous works [22–27], is based on successive approximations of the Hamiltonian using canonical transformations, and has been applied to the study of relative motion [28], frozen orbits [29], and mission design [30]. In addition, other families of perturbation methods have been used, for example Lindstedt-Poincaré and Krylov–Bogoliubov–Mitropolsky [31]. Although it has been the focus of less research, the $J_2$ problem with drag has also been treated using similar methods, namely the von Zeipel method [32], Krylov–Bogoliubov–Mitropolsky method [33], and simple power series expansion [34].

In contrast to mean elements, osculating elements provide a more straightforward description of the state of a satellite, but lead to longer equations when applying perturbation methods. As a consequence, analytical solutions to the $J_2$ problem based on osculating orbital elements are less common in the literature, and, to the authors' knowledge, do not exist for the $J_2$ problem with drag. In the $J_2$ problem, some examples include the application of the Lindstedt-Poincaré method with expansion in multiple frequencies [35, 36], simple power series expansion [37, 38], operator theory through the Koopman operator [8, 9], and Picard iterations [39].

In this work, we follow an approach based on osculating elements, with the goal of obtaining constant linear operators representing the dynamics in the $J_2$ problem and in the $J_2$ problem with drag, for any initial state, under the assumption of low-eccentricity orbits. To this end, we propose two methods based on the Lindstedt-Poincaré expansion to generate linear operators. The first aims to solve conservative systems, where the frequency of the solution does not change over time, which is the case of the $J_2$ problem. The second method is based on a modification of the Lindstedt-Poincaré expansion that allows the frequency of the solution to dynamically adapt to non-conservative dynamics, in this case applied to the $J_2$ problem with drag.

This paper is structured as follows. First, in Sec. II, we introduce the necessary background on perturbation theory and the Lindstedt-Poincaré method. Next, we describe the proposed method for generating an approximate linear operator representing the dynamics of a nonlinear system. This methodology is then applied to a simple toy problem, the Duffing oscillator, in Sec. III. The application to the more complex $J_2$ problem follows in Sec. IV. There, the used set of orbital elements is described, which allows writing the equations of motion in polynomial form. Then, the linear matrix representing this system is generated and applied to a low-eccentricity frozen orbit, and its performance is compared with previous definitions of the Koopman operator [8, 9]. The methodology for generating the linear operator is then applied to the $J_2$ problem with drag in Sec. V, using both the traditional Lindstedt-Poincaré method and the proposed modification to account for the changing frequency of the solution. The operators presented in this work are available at `https://github.com/MiguelAvillez/perturbation-theory-linear-operator`.

## II. Methodology

### A. Preliminaries

Most problems in astrodynamics do not have a closed-form analytical solution, examples of that being the zonal harmonics problem and the circular restricted three-body problem. However, in some cases, it is possible to obtain approximate analytical solutions using perturbation methods. One such method is the well-known Lindstedt-Poincaré method [40, 41], which we use in this work as the base for generating a linear operator.

The initial value problem can be described by the autonomous system of nonlinear differential equations

$$\begin{cases} \dfrac{d\boldsymbol{x}}{dt} = \boldsymbol{f}(\boldsymbol{x}; \varepsilon) \\ \boldsymbol{x}(t_0) = \boldsymbol{\eta} \end{cases} \quad (1)$$

where $\boldsymbol{x} \in \mathbb{R}^d$ represents the dependent variables (i.e. the state), $t \in \mathbb{R}$ is the independent variable, $\boldsymbol{\eta}$ are the initial conditions, $\boldsymbol{f}(\boldsymbol{x}; \varepsilon) : \mathbb{R}^d \times \mathbb{R} \to \mathbb{R}^d$ is a nonlinear function, and $\varepsilon$ is a small parameter ($\varepsilon \ll 1$). The Lindstedt-Poincaré method is based on computing the solution of $\boldsymbol{x}(t)$ as a power expansion in the small parameter $\varepsilon$

$$\boldsymbol{x}(t) := \sum_{i=0}^{n} \varepsilon^i \boldsymbol{x}_i(t) + \mathcal{O}\left(\varepsilon^{n+1}\right) \quad (2)$$

where $n$ is the order of the expansion, selected based on the desired accuracy of the approximate solution. The zeroth-order term $\boldsymbol{x}_0(t)$ is the solution of the problem with $\varepsilon = 0$, known as the unperturbed problem, which is required to have an analytical solution for a perturbation method to be integrable. Each $i^{th}$-order term $\boldsymbol{x}_i(t)$ for $i \geq 1$ is a correction to the unperturbed solution to account for the perturbation produced by $\varepsilon$, where the magnitude of the correction $\varepsilon^i \boldsymbol{x}_i(t)$ is smaller for higher values of $i$. For example, the first-order term $\boldsymbol{x}_1$ describes the corrections with magnitude $\mathcal{O}(\varepsilon)$, which correspond solely to the direct influence of $\varepsilon$ on each element of $\boldsymbol{x}$. The second-order term $\boldsymbol{x}_2$ describes corrections with magnitude $\mathcal{O}(\varepsilon^2)$, which includes smaller direct effects of $\varepsilon$ on each element of $\boldsymbol{x}$ and indirect effects induced by the effects of $\varepsilon$ on the other elements of $\boldsymbol{x}$.

Introducing the previous expansion into the left and right-hand sides of the initial value problem leads to a system of differential equations

$$\sum_{i=0}^{n} \varepsilon^i \frac{d\boldsymbol{x}_i(t)}{dt} \approx \sum_{i=0}^{n} \varepsilon^i \boldsymbol{f}_i(\boldsymbol{y}) \quad (3)$$

where $\boldsymbol{f}_i$ is a nonlinear function of order $\mathcal{O}(1)$, and $\boldsymbol{y} \in \mathbb{R}^{(n+1) \times d}$ is the extended state vector

$$\boldsymbol{y} := \left[\boldsymbol{x}_0^T, \boldsymbol{x}_1^T, \ldots, \boldsymbol{x}_n^T\right]^T \quad (4)$$

defined as the concatenation of all the $i^{th}$ order terms $\boldsymbol{x}_i$, with $i = 1, \ldots, n$. Identifying the coefficients with the same power of $\varepsilon$ (the same order), it is possible to construct the system of differential equations describing the evolution of the extended state vector

$$\frac{d\boldsymbol{y}}{dt} = \boldsymbol{g}(\boldsymbol{y}) \quad (5)$$

which no longer depends on the parameter $\varepsilon$. Using the initial conditions $\boldsymbol{x}_0(t_0) = \boldsymbol{x}(t_0)$ and $\boldsymbol{x}_i(t_0) = 0$ for $i = 1, \ldots, n$, it is possible to analytically solve this system by sequentially integrating the differential equations for each $\boldsymbol{x}_i(t)$, for $i = 0, ..., n$. The solution for $x(t)$ is then obtained by introducing the solutions for each $x_i(t)$ into Eq. (2) [40, 41].

Such an approximate solution $x(t)$, based on a simple power expansion, generally becomes inaccurate after a relatively short interval of the independent variable, due to the presence of secular terms in the approximate solution. To mitigate this issue, the Lindstedt-Poincaré method [40, 41] performs an additional expansion of the frequency $\omega$ of the solution

$$\omega \approx \sum_{i=0}^{n} \varepsilon^i \omega_i \qquad (6)$$

through the regularization

$$\tau = \omega t \qquad (7)$$

As before, introducing the expansions into the initial value problem will lead to a system in the same form of Eq. (3), which, after separating the terms based on the order of $\varepsilon$, can be solved sequentially for each $x_i(\tau)$, where the frequencies $\omega_i$ are selected to eliminate the secular terms in the solutions $x_i(\tau)$.

Both described perturbation methods are suited to solve systems of differential equations consisting of the sum of a linear part, which has a closed-form analytical solution, with a small nonlinear part. To apply these methods, the initial value problem has to be written such that the differential equations for the expanded dependent variables $x_i$, with $i = 0, ..., n$, can be integrated analytically. This is guaranteed to be the case if the zeroth order system is integrable and the differential equations resulting from the power expansion are in the form of polynomials or trigonometric polynomials.

## B. Generating a Linear Operator

We propose generating a linear operator representing the nonlinear system based on the application of the Lindstedt-Poincaré method. That is, the goal is to find a constant matrix $M$, independent of the state of the particle, that allows writing the system as

$$\begin{cases} \dfrac{d\boldsymbol{v}}{d\tau} = M\boldsymbol{v} \\ \boldsymbol{v}(\tau_0) = \boldsymbol{v}_0 \end{cases} \qquad (8)$$

for a set of basis functions $\boldsymbol{v}$ (their selection is described later) that expand the configuration space of the system, but are able to represent some of its nonlinearities. This provides the advantages associated with both operator theory and classical perturbation theory. In particular, the linear representation of the system enables the application of the numerous existing techniques for the analysis of linear systems, for example, to study their stability and control, as well as using the various methods for solving linear systems, for instance, through an eigendecomposition. From the side of perturbation theory, the proposed method maintains the typical advantages of the Lindstedt-Poincaré method, namely the clear physical meaning of the terms of the nonlinear dynamics being neglected and the long-term stability of the approximated solution.

As previously mentioned, it is only possible to apply the Lindstedt-Poincaré method to a perturbed problem if the system of differential equations resulting from the power expansion has an analytical solution, which is guaranteed to be the case if the expanded differential equations are either in the form of polynomials or trigonometric polynomials. Since the latter can always be written as the former (by defining the trigonometric functions as new variables), we focus on generating an expanded system in polynomial form. Note that this does not mean that the nonlinear differential equations need to be in polynomial form, it is simply necessary that the resultant expansion is polynomial.

Supposing that the expanded equations are polynomial, the differential equations resulting from the power expansion [Eq. (5)] are sums of monomials with the form

$$\frac{dx_{i,j}}{dt} = \sum_{k=1}^{q_{i,j}} C_k u_k \tag{9}$$

where $x_{i,j}$ represents the $j^{th}$ element of $x_i$, $C_k \in \mathbb{R}$ are constant coefficients, $q_{i,j}$ is the number of monomials in the polynomial, and $u_k$ is a monomial on the variables of the extended state $y$. To construct the linear operator matrix, we first define the set of basis functions $v$ to include the extended state vector $y$. The vector $v$ is then extended by defining all the monomials $u_k$ as new basis functions, and the row of $M$ representing $dx_{i,j}/dt$ is filled by placing the coefficients $C_k$ in the appropriate positions. This process is repeated for each monomial $u_k$, where again each new monomial appearing in the equation for $du_k/dt$ is defined as a new basis function. This whole procedure is executed for each element of the extended state vector $y$, producing an exact linear representation of the differential equations in the extended state $y$ [Eq. (5)], which in turn defines an approximate linear representation of the differential equations for the state $x$ [Eq. (1)].

*1. Representability*

Following this method, we find the vector of basis functions $v$ and the linear operator $M$. However, for the method to have practical use, it is necessary to ensure that $M$ is finite dimensional, i.e. that the process of defining each monomial as a basis function will not lead to the creation of infinite new monomials. Before stating the conditions for that to happen, consider the structure of the differential equations resulting from the power expansion. The zeroth order equations have the form

$$\frac{dx_{0,i}}{dt} = \sum_{k=1}^{q_{i,j}} C_k \prod_{j=1}^{d} x_{0,j}^{a_{k,0,j}} \tag{10}$$

where $a_{k,0,j} \in \mathbb{N}_0$ is an exponent, meaning that the zeroth order equations may be nonlinear, and $d$ is the number of dimensions (recall that $x \in \mathbb{R}^d$). Meanwhile, the $n^{th}$ order equations (with $n \geq 1$) can be represented by

$$\frac{dx_{n,i_1}}{dt} = \sum_{i_2=1}^{d} g_{i_1 i_2}(x_0) x_{n,i_2} + l_{i_1}(x_0, x_1, \ldots, x_{n-1}) \tag{11}$$

where $i_1 \in \{1, \ldots, d\}$. The $g_{i_1 i_2}(\boldsymbol{x}_0)$ function is a sum of monomials constituted by zeroth order variables, with $g_{i_1 i_2}(\boldsymbol{x}_0) x_{n,i_2}$ representing all the monomials which involve $n^{th}$ order variables. Note that, due to the used power expansion, the $n^{th}$ order equation is at most linear with respect to $n^{th}$ order variables (powers of $n^{th}$ order variables would have an order larger than $n$), and the $n^{th}$ order variables can only be multiplied by zeroth order terms (multiplication by terms of larger order would create monomials with an order larger than $n$). The $l_{i_1}(\boldsymbol{x}_0, \boldsymbol{x}_1, \ldots, \boldsymbol{x}_{n-1})$ function is a sum of monomials which collects all terms constituted by variables of order less than or equal to $n-1$. This separation into monomials that include and do not include $n^{th}$ order variables is used to determine the conditions under which a system can be represented in a finite way. To simplify the notation, the dependent variables of the $g$ and $l$ functions are omitted in the following paragraphs.

*Definition.* A monomial or collection of monomials is "representable" if it can be represented linearly by a finite-dimensional matrix $M$.

*Theorem* 1. Let the zeroth order differential equations be such that none of them depends on $x_{0,j}$, where $x_{0,j}$ are the state elements that have a nonlinear differential equation $dx_{0,j}/dt$. Furthermore, let the $n^{th}$ order differential equations ($n \geq 1$) be such that $g_{i_1 i_1}$, $g_{i_1 i_2} g_{i_2 i_1}$, $g_{i_1 i_2} g_{i_2 i_3} g_{i_3 i_1}$, $\ldots$, $g_{i_1 i_2} g_{i_2 i_3} \cdots g_{i_{d-1} i_d} g_{i_d i_1}$ are constants independent of any variable for $i_1, i_2, \ldots, i_d \in \{1, 2, \ldots, d\}$, and $i_2 \neq i_1$, $i_3 \notin \{i_1, i_2\}$, $\ldots$, $i_d \notin \{i_1, i_2, \ldots, i_{d-1}\}$. Then the system is representable.

*Proof.* The proof is done by induction.

*Base case:* Let $u$ be a general zeroth order monomial of the expanded system which, due to the polynomial structure of the differential equation, can be represented by

$$u = \prod_{j=1}^{d} x_{0,j}^{a_{0,j}} \tag{12}$$

with $a_{0,j} \in \mathbb{N}_0$. Let $p_u$ be the order of the monomial* $u$, that is, $p_u := \sum_{j=1}^{d} a_{0,j}$ is the sum of the exponents of $u$. Analyzing the derivative of $u$, we obtain

$$\frac{du}{dt} = \sum_{i=1}^{d} a_{0,i} \cdot x_{0,i}^{a_{0,i}-1} \frac{dx_{0,i}}{dt} \prod_{\substack{j=1 \\ j \neq i}}^{d} x_{0,j}^{a_{0,j}} \tag{13}$$

where the differential equation $dx_{0,i}/dt$ contains monomials with a maximum order of $q_i$. Therefore, analyzing the order of the $i^{th}$ term of the previous equation, we obtain

$$p_i = a_{0,i} - 1 + q_i + \sum_{\substack{j=1 \\ j \neq i}}^{d} a_{0,j} = \sum_{j=1}^{d} a_{0,j} + q_i - 1 = p_u + q_i - 1 \quad \text{if} \quad a_{0,i} \neq 0$$

$$p_i = 0 \quad \text{if} \quad a_{0,i} = 0 \tag{14}$$

---

*The order of the monomial is distinct from the order of the expansion; in the base-case section of the proof the word "order" is always used to refer to the order of the monomial, in the rest of the paper it is always used to refer to the order of the expansion.

If $p_i \leq p_u \; \forall i \in \{1, 2, \ldots, d\}$, then $u$ can be linearly represented by a finite ring of polynomials with order less than or equal to $p_u$. For $a_{0,i} = 0$ this is automatically satisfied, for $a_{0,i} \neq 0$ we have

$$p_i \leq p_u \Leftrightarrow p_u + q_i - 1 \leq p_u \Rightarrow q_i \in \{0, 1\} \tag{15}$$

Therefore, even if there is a nonlinear differential equation $dx_{0,i}/dt$, if $a_{0,i} = 0$ then the monomial $u$ is representable.

*Inductive hypothesis:* Assume that each monomial constituted by variables of order less than or equal to $n-1$ ($\prod_{i=0}^{n-1} \prod_{j=1}^{d} x_{i,j}^{a_{i,j}}$, with $a_{i,j} \in \mathbb{N}_0$) is representable.

*Inductive step:* To simplify the notation, the $n^{th}$ order terms are represented by $z_{i_1} := x_{n,i_1}$ with $i_1 \in \{1, \ldots, d\}$. Following Eq. (11), their derivatives are given by

$$\frac{dz_{i_1}}{dt} = g_{i_1 i_1}(\boldsymbol{x}_0) z_{i_1} + \sum_{\substack{i_2=1 \\ i_2 \neq i_1}}^{d} g_{i_1 i_2}(\boldsymbol{x}_0) z_{i_2} + l_{i_1}(\boldsymbol{x}_0, \boldsymbol{x}_1, \ldots, \boldsymbol{x}_{n-1}) \tag{16}$$

The dependent variables of the $g$ and $l$ functions are omitted in the following paragraphs. For these monomials to be representable, $dz_{i_1}/dt$ has to be at most linear with respect to $z_{i_1}$ (and so does any $m^{th}$ derivative of $z_i$), otherwise attempting to represent a monomial involving $z_{i_1}$ would lead to an increase of the exponents of the zeroth order terms multiplying $z_{i_1}$. Therefore, the term $g_{i_1 i_1} z_{i_1}$ is representable if and only if $g_{i_1 i_1}$ is a constant independent of any variable. The term $l_{i_1}$ is representable by the inductive hypothesis.

The terms $p_{i_1 i_2} := g_{i_1 i_2} z_{i_2}$ with $i_2 \neq i_1$ can now be analyzed by taking the derivative

$$\frac{dp_{i_1 i_2}}{dt} = g'_{i_1 i_2} z_{i_2} + g_{i_2 i_2} p_{i_1 i_2} + g_{i_1 i_2} g_{i_2 i_1} z_{i_1} + \sum_{\substack{i_3=1 \\ i_3 \neq i_1, i_2}}^{d} g_{i_1 i_2} g_{i_2 i_3} z_{i_3} + g_{i_1 i_2} l_{i_2} \tag{17}$$

where $g_{i_2 i_2} p_{i_1 i_2}$ is representable since $g_{i_2 i_2}$ is a constant (due to the previously found condition), $g_{i_1 i_2} l_{i_2}$ is representable by the inductive hypothesis, and $g_{i_1 i_2} g_{i_2 i_1} z_{i_1}$ is representable if and only if $g_{i_1 i_2} g_{i_2 i_1}$ is a constant independent of any variable for $i_2 \neq i_1$.

This process can be continued by taking the derivative of the newly formed terms $p_{i_1 i_2 i_3} := g_{i_1 i_2} g_{i_2 i_3} z_{i_3}$

$$\frac{dp_{i_1 i_2 i_3}}{dt} = (g_{i_1 i_2} g_{i_2 i_3})' z_{i_3} + g_{i_3 i_3} p_{i_1 i_2 i_3} + g_{i_2 i_3} g_{i_3 i_2} p_{i_1 i_2} + g_{i_1 i_2} g_{i_2 i_3} g_{i_3 i_1} z_{i_1} + \sum_{\substack{i_4=1 \\ i_4 \neq i_1, i_2, i_3}}^{d} g_{i_1 i_2} g_{i_2 i_3} g_{i_3 i_4} z_{i_4} + g_{i_1 i_2} g_{i_2 i_3} l_{i_3} \tag{18}$$

The terms $g_{i_3 i_3} p_{i_1 i_2 i_3}$ and $g_{i_2 i_3} g_{i_3 i_2} p_{i_1 i_2}$ are representable because $g_{i_3 i_3}$ and $g_{i_2 i_3} g_{i_3 i_2}$ are constants, $g_{i_1 i_2} g_{i_2 i_3} l_{i_3}$ is representable by the inductive hypothesis, and $g_{i_1 i_2} g_{i_2 i_3} g_{i_3 i_1} z_{i_1}$ is representable if and only if $g_{i_1 i_2} g_{i_2 i_3} g_{i_3 i_1}$ is a constant independent of any variable for $i_3 \notin \{i_1, i_2\}$. The process of defining each new monomial is then repeated for the subsequent indices, up to the number of dimensions of the system ($d$), eventually leading to $p_{i_1 i_2 \ldots i_d} :=$ $g_{i_1 i_2} g_{i_2 i_3} \cdots g_{i_{d-1} i_d} z_{i_d}$ with $i_d \notin \{i_1, i_2, \ldots, i_{d-1}\}$. Taking the derivative of this term

$$\begin{aligned}
\frac{dp_{i_1 i_2 \ldots i_d}}{dt} &= (g_{i_1 i_2} g_{i_2 i_3} \cdots g_{i_{d-1} i_d})' z_{i_d} + g_{i_d i_d} p_{i_1 i_2 \ldots i_d} + g_{i_{d-1} i_d} g_{i_{d-1} i_d} p_{i_1 i_2 \ldots i_{d-1}} + \cdots + g_{i_2 i_3} \cdots g_{i_d i_2} p_{i_1 i_2} \\
&+ g_{i_1 i_2} g_{i_2 i_3} \cdots g_{i_{d-1} i_d} g_{i_d i_1} z_{i_1} + \sum_{\substack{i_{d+1}=1 \\ i_{d+1} \neq i_1, i_2, \ldots, i_d}}^{d} g_{i_1 i_2} \cdots g_{i_d i_{d+1}} z_{i_{d+1}} + g_{i_1 i_2} g_{i_2 i_3} \cdots g_{i_{d-1} i_d} l_{i_d}
\end{aligned} \quad (19)$$

where the sum $\sum_{i_{d+1}=1,\, i_{d+1} \neq i_1, i_2, \ldots, i_d}^{d} g_{i_1 i_2} \cdots g_{i_d i_{d+1}} z_{i_{d+1}}$ is zero because there is no index $i_{d+1} \in \{1, \ldots, d\}$ that is able to satisfy the restriction $i_{d+1} \notin \{i_1, i_2, \ldots, i_d\}$. Similarly to the previous derivatives, the term $g_{i_1 i_2} g_{i_2 i_3} \cdots g_{i_{d-1} i_d} g_{i_d i_1} z_{i_1}$ is representable if and only if $g_{i_1 i_2} g_{i_2 i_3} \cdots g_{i_{d-1} i_d} g_{i_d i_1}$ is a constant independent of any variable for $i_d \notin \{i_1, i_2, \ldots, i_{d-1}\}$, the term $g_{i_1 i_2} g_{i_2 i_3} \cdots g_{i_{d-1} i_d} l_{i_d}$ is representable by the inductive hypothesis, and the remaining terms are constants multiplied by already defined terms.

The only terms not yet analyzed are $g'_{i_1 i_2} z_{i_2}$, $(g_{i_1 i_2} g_{i_2 i_3})' z_{i_3}$, ..., $(g_{i_1 i_2} g_{i_2 i_3} \cdots g_{i_{d-1} i_d})' z_{i_d}$. As discussed for the base case, the derivative of each $g$ function is composed by the elements of a ring of polynomials with maximum sum of exponents equal to the $g$ function itself, therefore allowing the finite representation of these terms.

Finally, since the $n^{th}$ order variables can be represented by a finite number of terms, monomials constituted by a collection of these and lower order terms are also representable. □

The determined conditions on the $n^{th}$ order differential equations ($n \geq 1$) can be interpreted as preventing the formation of cycles between the variable $z_i$ and itself over which the exponents of the zeroth order terms multiplying $z_i$ increase (cycles over which the exponents do not increase are allowed), as this would mean the formation of new higher-exponent monomials every time a monomial involving $z_i$ is represented. The existence of these cycles can also be visualized by representing the $n^{th}$ order ($n \geq 1$) differential equations as a weighted directed graph. First, represent each $n^{th}$ order variable $x_{n,j}$ by a node. Then, represent the monomials in $dx_{n,j}/dt$ that are linear with respect to $n^{th}$ order variables by weighted edges, where the weights are the $g$ functions. For example, for $n = 1$, $dx_1/dt = x_0 y_1 + x_0 y_0^2 + z_0 y_0 z_1$ would be represented by a directed edge between $x_1$ and $y_1$ with weight $x_0$, and a directed edge between $x_1$ and $z_1$ with weight $z_0 y_0$. If the directed graph contains any cycle over which there is an increase of the exponents of zeroth order terms, then the conditions of the theorem are not satisfied.

A similar set of conditions representing the absence of exponent-increasing cycles could also be derived for the zeroth order differential equations. When Theorem 1 is developed for the zeroth order system, it generates a more restrictive set of conditions than the ones required for representabillity, however, it also provides a simpler proof for zeroth order that covers all the examples analyzed in the following sections.

Note that Theorem 1 is formulated without any assumptions on the structure of the differential equations besides the ones related to the definition of the orders themselves. Therefore, the theorem is valid when the expanded differential equations are obtained using regularization (for instance, through the Lindstedt-Poincaré method). Alternatively, the conditions of this theorem can be simplified when applying it to simple power expansions without any regularization, through the following corollary.

*Corollary* 1.1. If the expanded differential equations result from a simple power expansion without regularization, and no zeroth order equation depends on the state elements $x_{0,j}$ that have a nonlinear differential equation $dx_{j,0}/dt$, then the system is representable.

*2. Implementation*

The procedure for constructing the operator matrix is implemented through Algorithm 1, where the monomials occurring in the differential equations are recursively defined as new basis functions. The pseudo-code uses 1-based array indexing represented by parenthesis. All the variables are assumed to be passed by reference, with each function being able to modify the passed variables. When allocating the $M$ matrix, we assume a fixed user-specified number of basis functions; at the end of the algorithm, $M$ should be truncated to remove the unused space if the specified number is larger than the true one.

This algorithm requires as input the length of the extended state, the maximum number of monomials in the differential equations, the predicted number of basis functions, and $F$, a three-dimensional array representing the equations of motion. $F$ has dimensions $(n+1)d \times \max q_{i,j} \times ((n+1)d+1)$, where $\max q_{i,j}$ is the maximum value of $q_{i,j}$ for all the differential equations (i.e. the length of the equation with the maximum number of monomials). The sub-array $F(i, j, :)$ represents the $j^{th}$ monomial of the $i^{th}$ differential equation. $F(i, j, 1)$ is the coefficient of this monomial and $F(i, j, k)$ with $k > 1$ is the exponent of the $(k-1)^{th}$ element of the extended state vector **y**. For example, assume a system with ordered variables $x, y, z$. If the $j^{th}$ monomial of the $i^{th}$ equation is $4x^3z$, it would be represented by $F(i, j, 1) = 4$, $F(i, j, 2) = 3$, $F(i, j, 3) = 0$, and $F(i, j, 4) = 1$.

The algorithm used to generate the operator matrix requires being able to identify each monomial (an array) by a unique key (a single number). This can be achieved through Algorithm 2, which is based on Ref. [42]. This algorithm counts the monomials based on combinatorics, with $\binom{a}{b}$ representing the binomial coefficient, using graded lexicographic order.

### III. Duffing Oscillator

To showcase the described methodology we first apply it to the Duffing oscillator. This is a very simple example, making it possible to describe all the required steps in detail. The Duffing oscillator can be described by the differential equations

$$\frac{dq}{dt} = p$$
$$\frac{dp}{dt} = -q - \varepsilon q^3 \qquad (20)$$

CREATEOPERATORMATRIX(*F, nDim, nTerms, nBasis*)
**Input:** *F* (derivatives), *nDim* (length of extended state), *nTerms* (maximum number of monomials in the differential equations), *nBasis* (number of basis functions)
**Output:** *M* (operator matrix), *mons* (monomials matrix)
    *mons* = *zeros*(*nBasis, nDim*)
    *M* = *zeros*(*nBasis, nBasis*)
    *sorting* = *zeros*(*nBasis*, 2)
    *counter* = 1
    **for** *i* = 1, ..., *nDims* **do**
        *mon* = *zeros*(1, *nDims*)
        *mon*(*i*) = 1
        PROCESSMONOMIAL(*mon, counter*)         // Process each basis function

PROCESSMONOMIAL(*mon, counter, F, nDim, nTerms, mons, M, sorting*)
**Input:** *mon* (monomial), *counter* (index of the current monomial), *F, nDim, nTerms, mons, M, sorting*
**Output:** *monRowM* (row of *M* associated with *mon*)
    *key* = MON2KEY(*mon, nDim*)
    *index* = BINARYSEARCH(*sorting*(:, 1), *key*)
    **if** *index* = *NAN* **then**     // Process monomial if it wasn't found in sorting
        *mons*(*counter*, :) = *mon*     // Save mon to mons
        *monRowM* = *counter*     // Row of M associated with mon
        SORTEDINSERT(*sorting*, [*key, counter*])
        *counter* += 1
        // Derivative of mon with respect to time
        **for** *k* = 1, ..., *nDim* **do**
            **if** *mon*(*k*) > 0 **then**
                **for** *l* = 1, ..., *nTerms* **do**
                    **if** *F*(*k, l*, 1) ≠ 0 **then**
                          // Time derivative due to basis function *x*: $\frac{\partial mon}{\partial x}\frac{dx}{dt}$
                        *newMon* = *mon*
                        *newMon*(*k*) −= 1
                        *newMonCoefficient* = *mon*(*k*) · *F*(*k, l*, 1)
                        **for** *p* = 1, ..., *nDim* **do**
                            *newMon*(*k*) += *F*(*k, l, p* + 1)
                        *newMonRowM* = PROCESSMONOMIAL(*newMon, counter, F, nDim, nTerms, mons, M, sorting*)
                        // Insert derivative in matrix
                        *M*(*monRowM, newMonRowM*) += *newMonCoefficient*

    **else**
        *monRowM* = *sorting*(*index*, 2)
    **return** *monRowM*

BINARYSEARCH(*vectorOfKeys, targetKey*)
    **if** *targetKey* in *vectorOfKeys* **then**
        **return** *index* of *targetKey*
    **else**
        **return** *NAN*

SORTEDINSERT(*sorting*, [*key, counter*])
    Insert the row [*key, counter*] into *sorting*, such that the first column of *sorting* remains sorted

**Algorithm 1:** Generation of the operator matrix.

```
Mon2Key(mon, nDim)
Input: mon (monomial), nDim (length of extended state)
Output: key (index of the monomial)
    // Determine the order of the current monomial
    monOr = 0
    for i = 1, ..., nDim do
     ⌊ monOr += mon(i)
    // Count monomials with order lower than the order of mon
    key = 0                              // Index of [0, 0, ..., 0] monomial
    for i = 0, ..., monOr − 1 do
     ⌊ key += (i + monOr − 1
                    i)
    // Count monomials with the same order as mon
    counter = 0
    for i = 1, ..., nDim − 1 do
        if mon(i) > 0 then
            for j = 0, ..., mon(i) − 1 do
             ⌊ key += ((nDim − 1 − i) + (monOr − counter)
                                nDim − 1 − i)
            counter += 1
    return key
```

**Algorithm 2:** Conversion of monomial to key.

where $q$ is the position, $p$ the velocity, $\varepsilon$ the small parameter, and $t$ the time. This system of differential equations has an analytical solution which can be obtained using elliptic integrals. Nevertheless, it is also possible to obtain an approximate solution using a Lindstedt-Poincaré expansion, in line with the methodology described in the previous section. To obtain a second order solution, the state variables are expanded as

$$q \approx q_0 + q_1\varepsilon + q_2\varepsilon^2$$
$$p \approx p_0 + p_1\varepsilon + p_2\varepsilon^2 \tag{21}$$

where the subscripts 0, 1, and 2 indicate, respectively, zeroth, first, and second order variables; the zeroth order variables correspond to the unperturbed problem. The frequency of the solution is controlled through a time regularization $\tau = \omega t$, with frequency

$$\omega = \omega_0 + \omega_1\varepsilon + \omega_2\varepsilon^2 \tag{22}$$

where $\omega_0 = 1$ corresponds to the unperturbed frequency. Introducing these expansions into Eq. (20) and separating the equations based on the order of the small parameter, we obtain the system of differential equations

$$\frac{dq_0}{d\tau} = p_0$$
$$\frac{dp_0}{d\tau} = -q_0$$

$$\frac{dq_1}{d\tau} = -\omega_1 p_0 + p_1$$
$$\frac{dp_1}{d\tau} = \omega_1 q_0 - q_0^3 - q_1$$
$$\frac{dq_2}{d\tau} = -\omega_2 p_0 - \omega_1(p_1 - \omega_1 p_0) + p_2$$
$$\frac{dp_2}{d\tau} = \omega_2 q_0 - \omega_1(\omega_1 q_0 - q_0^3 - q_1) - 3q_0^2 q_1 - q_2 \quad (23)$$

The frequencies $\omega_1$ and $\omega_2$ are determined by analytically solving these differential equations, and selected to ensure that the secular terms of the solution are zero, leading to

$$\omega_1 = \frac{3}{8}\left(q(t_0)^2 + p(t_0)^2\right)$$
$$\omega_2 = -\frac{3}{256}\left(7q(t_0)^4 + 46q(t_0)^2 p(t_0)^2 + 23p(t_0)^4\right) \quad (24)$$

where $q(t_0)$ and $p(t_0)$ are the initial conditions.

Having the system of expanded differential equations, it is now possible to look for the linear matrix representing them. To ensure that this matrix is independent of the initial conditions, $\omega_1$ and $\omega_2$ are defined as basis functions, therefore expanding the system in Eq. (23) with

$$\frac{d\omega_1}{d\tau} = \frac{d\omega_2}{d\tau} = 0 \quad (25)$$

which results in an extended state vector $[q_0, \ p_0, \ q_1, \ p_1, \ q_2, \ p_2, \ \omega_1, \ \omega_2]^T$. For applying Algorithm 1, the system of differential equations needs to be represented by an array $F$. For example, the first equation (for $q_0$) is represented by non-zero entries $F(1, 1, 1) = 1$ and $F(1, 1, 3) = 1$, and the fourth equation (for $p_1$) is represented by $F(4, 1, 1) = -1$, $F(4, 1, 2) = 3$, $F(4, 2, 1) = 1$, $F(4, 2, 2) = 1$, $F(4, 2, 8) = 1$, $F(4, 3, 1) = -1$, and $F(4, 3, 4) = 1$. Applying Algorithm 1, the expanded system of differential equations is represented exactly (i.e. without any further approximation) by a constant matrix $M$ with size $36 \times 36$ with non-zero coefficients listed in Table 1 and associated vector of basis functions

$$v = \begin{bmatrix} q_0 & p_0 & \omega_1 & \omega_2 & q_0\omega_1 & p_0\omega_1 & q_0\omega_2 & p_0\omega_2 & q_0^3 & q_0^2 p_0 & q_0 p_0^2 & p_0^3 \\ q_0\omega_1^2 & p_0\omega_1^2 & q_0^3\omega_1 & q_0^2 p_0\omega_1 & q_0 p_0^2\omega_1 & p_0^3\omega_1 & q_0^5 & q_0^4 p_0 & q_0^3 p_0^2 & q_0^2 p_0^3 & q_0 p_0^4 & p_0^5 \\ q_1 & p_1 & q_1\omega_1 & p_1\omega_1 & q_0^2 q_1 & q_0 p_0 q_1 & p_0^2 q_1 & q_0^2 p_1 & q_0 p_0 p_1 & p_0^2 p_1 & q_2 & p_2 \end{bmatrix}^T$$

The operator matrix $M$ does not depend on the value of the small parameter or the state, therefore, after being determined, it can be applied to any set of initial conditions. Being based on a perturbation method, the accuracy of the approximate solution can be controlled by either increasing or decreasing the order of the used power expansion, which in turn will change the size of the operator matrix. For instance, a first order expansion is associated with an $11 \times 11$ operator, a second order one with a 36 × 36 operator, and third order with a 101 × 101 operator.

**Table 1** Non-zero coefficients of the operator matrix M representing the second-order Lindstedt-Poincaré approximation of the Duffing oscillator.

| Row | Col. | Coeff. | Row | Col. | Coeff. | Row | Col. | Coeff. | Row | Col. | Coeff. |
|---|---|---|---|---|---|---|---|---|---|---|---|
| 1  | 2  | 1  | 18 | 17 | -3 | 28 | 15 | -1 | 33 | 32 | -1 |
| 2  | 1  | -1 | 19 | 20 | 5  | 28 | 27 | -1 | 33 | 34 | 1  |
| 5  | 6  | 1  | 20 | 19 | -1 | 29 | 16 | -1 | 34 | 17 | 1  |
| 6  | 5  | -1 | 20 | 21 | 4  | 29 | 30 | 2  | 34 | 21 | -1 |
| 7  | 8  | 1  | 21 | 20 | -2 | 29 | 32 | 1  | 34 | 31 | -1 |
| 8  | 7  | -1 | 21 | 22 | 3  | 30 | 17 | -1 | 34 | 33 | -2 |
| 9  | 10 | 3  | 22 | 21 | -3 | 30 | 29 | -1 | 35 | 8  | -1 |
| 10 | 9  | -1 | 22 | 23 | 2  | 30 | 31 | 1  | 35 | 14 | 1  |
| 10 | 11 | 2  | 23 | 22 | -4 | 30 | 33 | 1  | 35 | 28 | -1 |
| 11 | 10 | -2 | 23 | 24 | 1  | 31 | 18 | -1 | 35 | 36 | 1  |
| 11 | 12 | 1  | 24 | 23 | -5 | 31 | 30 | -2 | 36 | 7  | 1  |
| 12 | 11 | -3 | 25 | 6  | -1 | 31 | 34 | 1  | 36 | 13 | -1 |
| 13 | 14 | 1  | 25 | 26 | 1  | 32 | 15 | 1  | 36 | 15 | 1  |
| 14 | 13 | -1 | 26 | 5  | 1  | 32 | 19 | -1 | 36 | 27 | 1  |
| 15 | 16 | 3  | 26 | 9  | -1 | 32 | 29 | -1 | 36 | 29 | -3 |
| 16 | 15 | -1 | 26 | 25 | -1 | 32 | 33 | 2  | 36 | 35 | -1 |
| 16 | 17 | 2  | 27 | 14 | -1 | 33 | 16 | 1  |    |    |    |
| 17 | 16 | -2 | 27 | 28 | 1  | 33 | 20 | -1 |    |    |    |
| 17 | 18 | 1  | 28 | 13 | 1  | 33 | 30 | -1 |    |    |    |

We test the generated second-order operator using initial position $q(0) = 1$, initial velocity $p(0) = 0$, and small parameter $\varepsilon = 0.01$. The evolution of the state and position error over one and fifteen periods are plotted in Fig. 1, where the error is measured with respect to the numerical integration of the equations of motion using a Runge-Kutta-Dormand-Prince 5(4) method with absolute and relative tolerances of $10^{-13}$. Over one revolution, a maximum position error of $10^{-7}$ can be observed, consistent with a second order solution. Over a longer propagation of fifteen revolutions, the error grows in an oscillating manner, due to the discrepancy between the approximated perturbed frequency and the true one, and due to the propagation of the error associated with any approximated solution.

The same methodology to generate the linear operator can be applied using a simple power expansion of the Duffing oscillator, i.e. without executing any time regularization (which corresponds to setting $\omega = 1$). This approach produces a 22 × 22 matrix, without the need to analytically solve the expanded differential equations in order to find the perturbed frequencies. Over one revolution, this simplified solution has a similar error to the Lindstedt-Poincaré solution. However, it has a faster error growth over long-term propagations, resulting from the mismatch between the perturbed and unperturbed frequencies of the system. As such, this simplified approach is useful when focusing on short-term propagations.

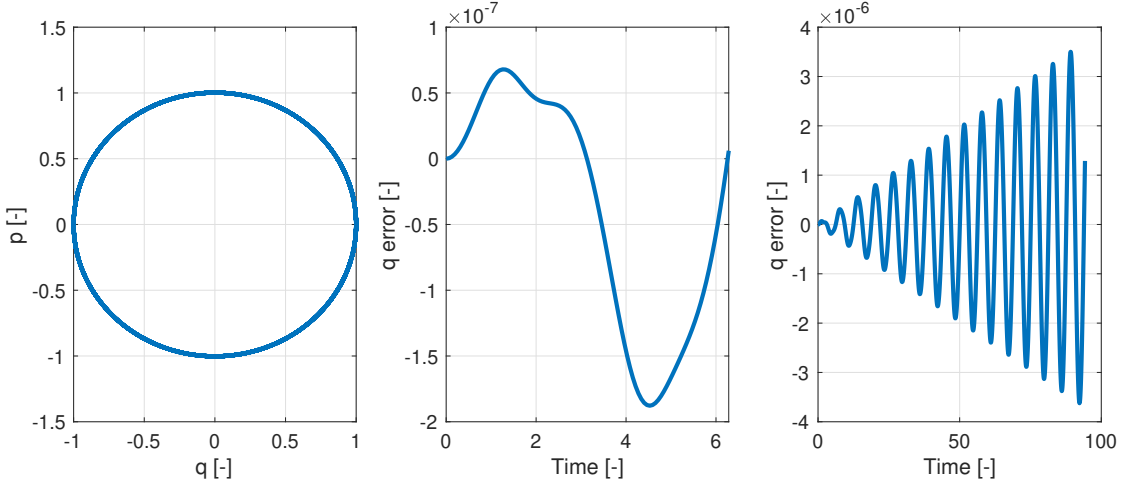

**Fig. 1 State space, and position error over one period and fifteen periods for the Duffing oscillator**

## IV. Orbit Under J$_2$ Perturbation

In this section we apply the proposed methodology for the generation of a linear operator to a more complex problem, the orbit of a satellite under the $J_2$ perturbation. The orbital elements used in this work are introduced and their polynomial differential equations are presented. Then, the linear operator that approximates these equations is generated and applied to an example orbit.

### A. Equations of Motion

*1. Equations of Motion in Spherical Coordinates*

The motion of a particle in a central gravity field with gravitational parameter $\mu$ and subject to perturbing accelerations $(a_r, a_\phi, a_\lambda)$ is described in spherical coordinates by

$$\begin{aligned}
\frac{dr}{dt} &= \dot{r} \\
\frac{d\dot{r}}{dt} &= -\frac{\mu}{r^2} + r\dot{\phi}^2 + r\dot{\lambda}^2 \cos^2\phi + a_r \\
\frac{d\phi}{dt} &= \dot{\phi} \\
\frac{d\dot{\phi}}{dt} &= -2\frac{\dot{\phi}\dot{r}}{r} - \dot{\lambda}^2 \sin\phi \cos\phi + \frac{a_\phi}{r} \\
\frac{d\lambda}{dt} &= \dot{\lambda} \\
\frac{d\dot{\lambda}}{dt} &= -2\frac{\dot{\lambda}\dot{r}}{r} + 2\dot{\lambda}\dot{\phi}\tan\phi + \frac{a_\lambda}{r\cos\phi}
\end{aligned} \qquad (26)$$

where $r$ represents the radial distance to the center of the celestial body, $\phi$ the latitude, and $\lambda$ the longitude of the orbiting particle with respect to the selected inertial frame.

*2. Variable Transformation*

To transform the equations of motion into a linear operator matrix, the equations need to be written in polynomial form, in order to guarantee that the perturbation expansion is integrable. To do so, a modification of the orbital elements proposed by Arnas [37] is used, consisting of the elements $(\beta, e_x, e_y, p, \Omega, C_\theta, S_\theta)$. These elements are defined based on the Keplerian orbital elements: semi-major axis ($a$), eccentricity ($e$), inclination ($i$), argument of periapsis ($\omega_p$), right ascension of the ascending node ($\Omega$), and argument of latitude ($\theta$). The variable $\beta$ is related to the inverse of the angular momentum

$$\beta = \left(\frac{R}{a(1-e^2)}\right)^{1/2} \tag{27}$$

where $R$ is the mean equatorial radius of the main celestial body (associated with the $J_2$ coefficient). The $e_x$ and $e_y$ variables are the two components of the eccentricity vector, along the line of apsides and perpendicular to it

$$e_x = e \cos \omega_p$$
$$e_y = e \sin \omega_p \tag{28}$$

The variable $p$ is the normalized product of the cosine of the inclination and the angular momentum

$$p = \cos i \left(\frac{a(1-e^2)}{R}\right)^{1/2} \tag{29}$$

This coincides with the normalized conjugate momenta of the longitude in the Hamiltonian representation of the zonal harmonics problem, which is a constant of motion [36, 43]. Finally, we define $C_\theta = \cos \theta$ and $S_\theta = \sin \theta$. These are used instead of $\theta$ to ensure that the equations of motion are polynomials instead of trigonometric polynomials; both $C_\theta$ and $S_\theta$ are necessary to disambiguate the sign of $\theta$. To obtain the differential equations, the following transformations between orbital elements $(\beta, e_x, e_y, p, \Omega, \theta)$ and spherical coordinates $(r, \dot{r}, \phi, \dot{\phi}, \lambda, \dot{\lambda})$ are used

$$\frac{1}{r} = \frac{\gamma}{h^2}\mu$$

$$\dot{r} = \frac{\mu}{h}(e_x \sin \theta - e_y \cos \theta)$$

$$\sin \phi = \sqrt{1 - p^2\beta^2} \sin \theta$$

$$\dot{\phi} = \frac{\sqrt{1 - p^2\beta^2} \cos \theta}{\cos \phi} \frac{h}{r^2}$$

$$\lambda = \begin{cases} \Omega + \arcsin \dfrac{\sin \theta p \beta}{\sqrt{1 - \sin^2 \theta(1 - p^2\beta^2)}} & \text{if } \cos \theta \geq 0 \\ \Omega - \arcsin \dfrac{\sin \theta p \beta}{\sqrt{1 - \sin^2 \theta(1 - p^2\beta^2)}} + \pi & \text{if } \cos \theta < 0 \end{cases}$$

$$\dot{\lambda} = \frac{hp\beta}{r^2 \cos^2 \phi} \tag{30}$$

with

$$h = \frac{\sqrt{\mu R}}{\beta^2} \tag{31}$$

$$\gamma = 1 + e_x \cos\theta + e_y \sin\theta \tag{32}$$

where $h$ is the magnitude of the orbital angular momentum. The inverse transformation, between spherical coordinates $(r, \dot{r}, \phi, \dot{\phi}, \lambda, \dot{\lambda})$ and orbital elements $(\beta, e_x, e_y, p, \Omega, \theta)$, is given by

$$\begin{aligned}
\beta &= \frac{\sqrt{\mu R}}{r^2 \sqrt{\dot{\phi}^2 + \dot{\lambda}^2 \cos^2\phi}} \\
e_x &= \left(\frac{h^2}{\mu r} - 1\right) \cos\theta + \frac{h\dot{r}}{\mu} \sin\theta \\
e_y &= \left(\frac{h^2}{\mu r} - 1\right) \sin\theta - \frac{h\dot{r}}{\mu} \cos\theta \\
p &= \frac{r^2 \dot{\lambda} \cos^2\phi}{\sqrt{\mu R}} \\
\Omega &= \begin{cases} \lambda - \arcsin\left(\dot{\lambda}\sin\phi \sqrt{\frac{\cos^2\phi}{\dot{\phi}^2 + \dot{\lambda}^2 \cos^2\phi \sin^2\phi}}\right) & \text{if } \cos\phi\dot{\phi} \geq 0 \\ \lambda + \arcsin\left(\dot{\lambda}\sin\phi \sqrt{\frac{\cos^2\phi}{\dot{\phi}^2 + \dot{\lambda}^2 \cos^2\phi \sin^2\phi}}\right) + \pi & \text{if } \cos\phi\dot{\phi} < 0 \end{cases} \\
\theta &= \begin{cases} \arcsin\left(\sin\phi \sqrt{\frac{\dot{\phi}^2 + \dot{\lambda}^2 \cos^2\phi}{\dot{\phi}^2 + \dot{\lambda}^2 \cos^2\phi \sin^2\phi}}\right) & \text{if } \cos\phi\dot{\phi} \geq 0 \\ -\arcsin\left(\sin\phi \sqrt{\frac{\dot{\phi}^2 + \dot{\lambda}^2 \cos^2\phi}{\dot{\phi}^2 + \dot{\lambda}^2 \cos^2\phi \sin^2\phi}}\right) + \pi & \text{if } \cos\phi\dot{\phi} < 0 \end{cases}
\end{aligned} \tag{33}$$

In this work we focus on low-eccentricity orbits, in particular assuming small eccentricities $e_x$ and $e_y$ in the order of $J_2$ (e.g. near-circular frozen orbits). As such, we introduce an additional change of variables [38]

$$\begin{aligned} X &= \frac{e_x}{J_2} \\ Y &= \frac{e_y}{J_2} \end{aligned} \tag{34}$$

This ensures that, like the other orbital elements, $X$ and $Y$ are in the order of unity for small-eccentricity orbits, which will later be useful to simplify the equations resulting from the perturbation method.

## 3. Equations of Motion

Having defined the used orbital elements and the associated transformations, we finally obtain the system of seven differential equations

$$\frac{d\beta}{dt} = \frac{R^{1/2}}{\mu^{1/2}\left(1 - J_2^2\left(X^2 + Y^2\right)\right)}\left(\gamma - 2 + J_2^2(XS_\theta - YC_\theta)^2\frac{1}{\gamma}\right)a_f$$

$$\frac{dX}{dt} = \frac{R^{1/2}}{J_2\mu^{1/2}\beta}\left(S_\theta a_r + \left(2C_\theta + J_2 S_\theta(XS_\theta - YC_\theta)\frac{1}{\gamma}\right)a_f + J_2\frac{S_\theta p \beta Y}{\gamma\sqrt{1 - p^2\beta^2}}a_h\right)$$

$$\frac{dY}{dt} = \frac{R^{1/2}}{J_2\mu^{1/2}\beta}\left(-C_\theta a_r + \left(2S_\theta - J_2 C_\theta(XS_\theta - YC_\theta)\frac{1}{\gamma}\right)a_f - J_2\frac{S_\theta p \beta X}{\gamma\sqrt{1 - p^2\beta^2}}a_h\right)$$

$$\frac{dp}{dt} = -\frac{R^{1/2}}{\mu^{1/2}\beta^2\gamma}C_\theta\sqrt{1 - p^2\beta^2}\,a_h - \frac{R^{1/2}p}{\mu^{1/2}\beta\left(1 - J_2^2\left(X^2 + Y^2\right)\right)}\left(\gamma - 2 + J_2^2(XS_\theta - YC_\theta)^2\frac{1}{\gamma}\right)a_f$$

$$\frac{d\Omega}{dt} = \frac{R^{1/2}}{\mu^{1/2}\beta\gamma\sqrt{1 - p^2\beta^2}}S_\theta a_h$$

$$\frac{dC_\theta}{dt} = -\frac{\mu^{1/2}}{R^{3/2}}S_\theta\beta^3\gamma^2\left(1 - \frac{R^2}{\mu}\frac{S_\theta p}{\beta^3\gamma^3\sqrt{1 - p^2\beta^2}}a_h\right)$$

$$\frac{dS_\theta}{dt} = \frac{\mu^{1/2}}{R^{3/2}}C_\theta\beta^3\gamma^2\left(1 - \frac{R^2}{\mu}\frac{S_\theta p}{\beta^3\gamma^3\sqrt{1 - p^2\beta^2}}a_h\right) \quad (35)$$

where $a_h$ is the acceleration in the direction of the angular momentum vector, $a_r$ the acceleration in the direction of the position vector, and $a_f$ the acceleration in the direction forming a right-handed frame with the previous vectors (these accelerations are specified with respect to a satellite-based frame).

When the perturbing acceleration is the $J_2$ term of the gravitational spherical harmonics, the previous system of differential equations becomes

$$\frac{d\beta}{dt} = 3\frac{J_2\mu^{1/2}}{R^{3/2}}\beta^8\gamma^3 S_\theta C_\theta\left(1 - \beta^2 p^2\right)$$

$$\frac{dX}{dt} = \frac{3}{2}\frac{\mu^{1/2}}{R^{3/2}}\beta^7\gamma^3 S_\theta\left(-C_\theta\left(1 - \beta^2 p^2\right)\left(4C_\theta + J_2 X\left(C_\theta^2 - S_\theta^2\right) + 2J_2 YC_\theta S_\theta + 3J_2 X\right)\right.$$
$$\left. - 2J_2\beta^2 p^2 YS_\theta + \gamma\left(3\left(1 - \beta^2 p^2\right)S_\theta^2 - 1\right)\right)$$

$$\frac{dY}{dt} = -\frac{3}{2}\frac{\mu^{1/2}}{R^{3/2}}\beta^7\gamma^3\left(J_2 XC_\theta^2\left(5\left(1 - \beta^2 p^2\right)S_\theta^2 - 1\right) + 2J_2 YC_\theta^3\left(1 - \beta^2 p^2\right)S_\theta\right.$$
$$\left. + C_\theta\left(J_2 YS_\theta + 1\right)\left(7\left(1 - \beta^2 p^2\right)S_\theta^2 - 1\right) - 2J_2\beta^2 p^2 XS_\theta^2\right)$$

$$\frac{dp}{dt} = 0$$

$$\frac{d\Omega}{dt} = -3\frac{J_2\mu^{1/2}}{R^{3/2}}\beta^8\gamma^3 p S_\theta^2$$

$$\frac{dC_\theta}{dt} = -\frac{\mu^{1/2}}{R^{3/2}}S_\theta\beta^3\gamma^2\left(3J_2\beta^6\gamma p^2 S_\theta^2 + 1\right)$$

$$\frac{dS_\theta}{dt} = \frac{\mu^{1/2}}{R^{3/2}}C_\theta\beta^3\gamma^2\left(3J_2\beta^6\gamma p^2 S_\theta^2 + 1\right) \quad (36)$$

Note that these equations are exact, i.e. no approximation has been made, and that they are completely polynomial, which allows applying the proposed method for generating a linear operator.

**B. Perturbation Method Without Control in Frequency**

*1. Construction of the Linear Operator*

Similarly to Ref. [37], a second-order solution to the equations of motion can be obtained by expanding the orbital elements according to a power series with small parameter $J_2$

$$\beta \approx \beta_0 + \beta_1 J_2 + \beta_2 J_2^2$$
$$X \approx X_1 + X_2 J_2$$
$$Y \approx Y_1 + Y_2 J_2$$
$$\Omega \approx \Omega_0 + \Omega_1 J_2 + \Omega_2 J_2^2$$
$$C_\theta \approx C_{\theta,0} + C_{\theta,1} J_2 + C_{\theta,2} J_2^2$$
$$S_\theta \approx S_{\theta,0} + S_{\theta,1} J_2 + S_{\theta,2} J_2^2 \tag{37}$$

This expansion assumes small eccentricities $e_x$ and $e_y$ in the order of $J_2$, such that $X$ and $Y$ are in the order of unity. Note that $X$ and $Y$ are only expanded up to the first order in $J_2$ because $X_2 J_2$ and $Y_2 J_2$ already correspond to second-order terms for $e_x$ and $e_y$ ($e_x J_2^2$ and $e_y J_2^2$). Additionally, since $p$ is a constant of motion it is not expanded as a power series.

Applying these expansions to the equations of motion and separating them based on the power of $J_2$ leads to the system of zeroth-order equations

$$\frac{d\beta_0}{dt} = 0$$
$$\frac{dp}{dt} = 0$$
$$\frac{d\Omega_0}{dt} = 0$$
$$\frac{dC_{\theta,0}}{dt} = -\omega_0 S_{\theta,0}$$
$$\frac{dS_{\theta,0}}{dt} = \omega_0 C_{\theta,0} \tag{38}$$

The zeroth-order equations describe the unperturbed system (i.e. Keplerian motion), hence the variables $\beta_0$ and $\Omega_0$ (associated, respectively, with the angular momentum and right ascension of the ascending node) are constant; $p$ is a constant of motion. Since these variables are constant, we have $\beta_0(t) = \beta(t_0)$, $p(t) = p(t_0)$, and $\Omega_0(t) = \Omega(t_0)$, with $t_0$ representing the initial time. The remaining variables have initial conditions $C_{\theta,0}(t_0) = C_\theta(t_0)$ and $S_{\theta,0}(t_0) = S_\theta(t_0)$.

The variable $\omega_0$ is the unperturbed frequency of the orbit

$$\omega_0 = \frac{\mu^{1/2}\beta_0^3}{R^{3/2}} \qquad (39)$$

which is very close to the unperturbed mean motion for small-eccentricity orbits.

In the same way, the system of first-order equations is obtained

$$\begin{aligned}
\frac{d\beta_1}{dt} &= -3\omega_0\beta_0^5\left(\beta_0^2 p^2 - 1\right)C_{\theta,0}S_{\theta,0} \\
\frac{dX_1}{dt} &= -\frac{3}{2}\omega_0\beta_0^4 S_{\theta,0}\left(\left(4 - 4\beta_0^2 p^2\right)C_{\theta,0}^2 + 3\left(\beta_0^2 p^2 - 1\right)S_{\theta,0}^2 + 1\right) \\
\frac{dY_1}{dt} &= \frac{3}{2}\omega_0\beta_0^4 C_{\theta,0}\left(7\left(\beta_0^2 p^2 - 1\right)S_{\theta,0}^2 + 1\right) \\
\frac{d\Omega_1}{dt} &= -3\omega_0\beta_0^5 p S_{\theta,0}^2 \\
\frac{dC_{\theta,1}}{dt} &= -\omega_0\left(S_{\theta,0}\left(3\beta_0^6 p^2 S_{\theta,0}^2 + 2X_1 C_{\theta,0} + 2Y_1 S_{\theta,0} + 3\frac{\beta_1}{\beta_0}\right) + S_{\theta,1}\right) \\
\frac{dS_{\theta,1}}{dt} &= \omega_0\left(C_{\theta,0}\left(3\beta_0^6 p^2 S_{\theta,0}^2 + 2X_1 C_{\theta,0} + 2Y_1 S_{\theta,0} + 3\frac{\beta_1}{\beta_0}\right) + C_{\theta,1}\right)
\end{aligned} \qquad (40)$$

with initial conditions $\beta_1(t_0) = \Omega_1(t_0) = C_{\theta,1}(t_0) = S_{\theta,1}(t_0) = 0$, $X_1(t_0) = X(t_0)$, and $Y_1(t_0) = Y(t_0)$. Finally, the system of second-order differential equations is

$$\begin{aligned}
\frac{d\beta_2}{dt} &= 3\omega_0\beta_0^4\left(2\beta_1\left(4 - 5\beta_0^2 p^2\right)C_{\theta,0}S_{\theta,0} \right. \\
&\quad \left. - \beta_0\left(\beta_0^2 p^2 - 1\right)\left(C_{\theta,0}\left(3S_{\theta,0}\left(X_1 C_{\theta,0} + Y_1 S_{\theta,0}\right) + S_{\theta,1}\right) + C_{\theta,1}S_{\theta,0}\right)\right) \\
\frac{dX_2}{dt} &= -\frac{3}{2}\omega_0\beta_0^3\left(\beta_0\left(-13X_1\left(\beta_0^2 p^2 - 1\right)C_{\theta,0}^3 S_{\theta,0} + C_{\theta,0}S_{\theta,0}\left(\left(8 - 8\beta_0^2 p^2\right)C_{\theta,1}\right.\right.\right. \\
&\quad \left.+ X_1\left(\beta_0^2 p^2\left(13 S_{\theta,0}^2 - 3\right) - 13 S_{\theta,0}^2 + 7\right)\right) - 2\left(\beta_0^2 p^2 - 1\right)C_{\theta,0}^2\left(7Y_1 S_{\theta,0}^2 + 2S_{\theta,1}\right) \\
&\quad + S_{\theta,0}^2\left(2Y_1\left(\beta_0^2 p^2\left(6S_{\theta,0}^2 + 1\right) - 6S_{\theta,0}^2 + 2\right) + 9\left(\beta_0^2 p^2 - 1\right)S_{\theta,1}\right) + S_{\theta,1}\Big) \\
&\quad \left.+ \beta_1 S_{\theta,0}\left(4\left(7 - 9\beta_0^2 p^2\right)C_{\theta,0}^2 + 3\left(9\beta_0^2 p^2 - 7\right)S_{\theta,0}^2 + 7\right)\right) \\
\frac{dY_2}{dt} &= \frac{3}{2}\omega_0\beta_0^3\left(\beta_0\left(2\left(X_1 C_{\theta,0}^2\left(13\left(\beta_0^2 p^2 - 1\right)S_{\theta,0}^2 + 2\right) + Y_1\left(\beta_0^2 p^2 - 1\right)C_{\theta,0}^3 S_{\theta,0}\right.\right.\right. \\
&\quad \left.+ C_{\theta,0}S_{\theta,0}\left(2Y_1\left(7\left(\beta_0^2 p^2 - 1\right)S_{\theta,0}^2 + 1\right) + 7\left(\beta_0^2 p^2 - 1\right)S_{\theta,1}\right) + \beta_0^2 p^2 X_1 S_{\theta,0}^2\right) \\
&\quad \left.\left.+ 7\left(\beta_0^2 p^2 - 1\right)C_{\theta,1}S_{\theta,0}^2 + C_{\theta,1}\right) + 7\beta_1 C_{\theta,0}\left(\left(9\beta_0^2 p^2 - 7\right)S_{\theta,0}^2 + 1\right)\right) \\
\frac{d\Omega_2}{dt} &= -3\omega_0\beta_0^4 p S_{\theta,0}\left(\beta_0\left(3 S_{\theta,0}\left(X_1 C_{\theta,0} + Y_1 S_{\theta,0}\right) + 2 S_{\theta,1}\right) + 8\beta_1 S_{\theta,0}\right) \\
\frac{dC_{\theta,2}}{dt} &= -\omega_0\left(3\beta_1\left(9\beta_0^5 p^2 S_{\theta,0}^3 + \frac{1}{\beta_0}\left(2 S_{\theta,0}\left(X_1 C_{\theta,0} + Y_1 S_{\theta,0}\right) + S_{\theta,1}\right)\right)S_{\theta,0}^3 + 3\frac{\beta_1^2}{\beta_0^2}S_{\theta,0}\right. \\
&\quad \left.+ 3\frac{\beta_2}{\beta_0}S_{\theta,0} + 9\beta_0^6 p^2 X_1 C_{\theta,0} + 9\beta_0^6 p^2 Y_1 S_{\theta,0}^4 + 9\beta_0^6 p^2 S_{\theta,0}^2 S_{\theta,1} + 2X_1 Y_1 C_{\theta,0}S_{\theta,0}^2 + X_1^2 C_{\theta,0}^2 S_{\theta,0}\right.
\end{aligned}$$

$$+ 2X_2 C_{\theta,0} S_{\theta,0} + 2X_1 C_{\theta,1} S_{\theta,0} + 2X_1 C_{\theta,0} S_{\theta,1} + Y_1^2 S_{\theta,0}^3 + 2Y_2 S_{\theta,0}^2 + 4Y_1 S_{\theta,0} S_{\theta,1} + S_{\theta,2}\Big)$$

$$\frac{dS_{\theta,2}}{dt} = \omega_0 \Big(3\beta_1 \Big(9\beta_0^5 p^2 C_{\theta,0} S_{\theta,0}^2 + \frac{1}{\beta_0}\left(2C_{\theta,0}\left(X_1 C_{\theta,0} + Y_1 S_{\theta,0}\right) + C_{\theta,1}\right)\Big) + 3\frac{\beta_1^2}{\beta_0} C_{\theta,0} + 3\frac{\beta_2}{\beta_0} C_{\theta,0}$$

$$+ 9\beta_0^6 p^2 X_1 C_{\theta,0}^2 S_{\theta,0}^2 + 9\beta_0^6 p^2 Y_1 C_{\theta,0} S_{\theta,0}^3 + 3\beta_0^6 p^2 C_{\theta,1} S_{\theta,0}^2 + 6\beta_0^6 p^2 C_{\theta,0} S_{\theta,0} S_{\theta,1} + 2X_1 Y_1 C_{\theta,0}^2 S_{\theta,0}$$

$$+ Y_1^2 C_{\theta,0} S_{\theta,0}^2 + 2Y_2 C_{\theta,0} S_{\theta,0} + 2Y_1 C_{\theta,1} S_{\theta,0} + 2Y_1 C_{\theta,0} S_{\theta,1} + X_1^2 C_{\theta,0}^3 + 2X_2 C_{\theta,0}^2 + 4X_1 C_{\theta,0} C_{\theta,1} + C_{\theta,2}\Big) \quad (41)$$

with initial conditions $\beta_2(t_0) = X_2(t_0) = Y_2(t_0) = \Omega_2(t_0) = C_{\theta,2}(t_0) = S_{\theta,2}(t_0) = 0$.

Having obtained systems of zeroth, first, and second order polynomial differential equations, and before we use them to generate the linear operator, it is necessary to rewrite them to satisfy the conditions of Theorem 1. In particular, no zeroth-order equation should depend on zeroth-order variables with a nonlinear differential equation, which is not the case in Eq. (38) due to the equations $dC_{\theta,0}/dt = -\omega_0 S_{\theta,0}$ and $dS_{\theta,0}/dt = \omega_0 C_{\theta,0}$. This can be addressed by performing a time regularization

$$\tau = \omega_0 t \quad (42)$$

Observe that the time $t$ depends linearly on $\tau$, as $\omega_0$ is a constant that only depends on the initial conditions. An additional condition for generating the linear operator is that the differential equations resulting from the power expansion should be in polynomial form. However, the performed regularization leads to the appearance of monomials multiplied by $1/\beta_0$ and $1/\beta_0^2$, which are not polynomial. These can be transformed into polynomial terms by defining the auxiliary variable $k_\beta = 1/\beta_0$, with time derivative

$$\frac{dk_\beta}{dt} = 0 \quad (43)$$

Now having systems of zeroth, first, and second order polynomial differential equations that satisfy the conditions of Theorem 1, it is possible to apply Algorithm 1 to determine the matrix $M$ that describes the system as $dv/d\tau = Mv$. The computation of $M$ is virtually instantaneous, taking an average of 0.15 s on a single-thread single-core MATLAB program, run on an Intel i7 2.6 GHz, 32 GB of RAM, macOS 13.6.1. The linear operator representing the expansion up to second order is a sparse $568 \times 568$ matrix. Meanwhile, considering only the expansion up to first order results in a $59 \times 59$ matrix. As previously mentioned, the size of the matrix and accuracy of the approximation are related to the order of the used power expansion. Using a higher-order expansion will produce a more accurate solution, but will also require the definition of additional basis functions, thus increasing the size of the matrix. Finally, note that the operator matrix is independent of the state, hence it only needs to be determined once and can afterwards be applied to any initial condition.

It is worth mentioning that the set of elements $(\beta, e_x, e_y, p, \Omega, \theta)$ is selected to minimize the size of the operator matrix $M$ when considering the $J_2$ perturbation. Other element sets are possible, for example $(\Lambda, e_x, e_y, C_i, \Omega, C_\theta, S_\theta)$

with

$$\Lambda = \left(\frac{R}{a}\right)^{1/2} \quad (44)$$

and $C_i = \cos i$. This set of elements leads to a $648 \times 648$ matrix for a second-order approximation, or $63 \times 63$ for a first-order approximation. With these elements, the equations of motion are not polynomial, but do become polynomial when doing a power expansion with the assumption of small eccentricities. The set of elements that leads to the smallest $M$ matrix is directly related to what perturbations are considered in the equations of motion; for example, using $p$ might not be beneficial when including perturbations other than the zonal harmonics, as in that case $p$ is no longer a constant of motion. This creates a large variety of potential approaches for defining the most appropriate orbital elements depending on the problem being considered.

*2. Application*

To show the performance of this method, the $568 \times 568$ operator matrix obtained from the second-order expansion is applied to a low-eccentricity frozen sun-synchronous orbit, with initial osculating elements $a$ = 7077.722 km, $e_x = 4.5742 \times 10^{-4}$, $e_y = 0$, $i$ = 98.186 deg, $\Omega$ = 42.0 deg, and $\theta$ = 0 deg [the frozen-orbit conditions are obtained using Eq. (49)]. The solution generated by the operator matrix is compared with the numerical integration of the equations of motion in spherical coordinates, using a Runge-Kutta 9(8) integrator [44] with absolute and relative tolerances of $10^{-13}$. When defining the propagation length, the orbital period is taken to correspond to the second-order solution of the nodal period determined in Ref. [38].

The evolution of the Keplerian orbital elements over one revolution (1.645 hour) is represented in Fig. 2. Fig. 3 shows the error of the second-order analytical solution with respect to the numerical propagation. The analytical solution can be observed to be very accurate, with a maximum semi-major axis error of 0.4 m. It can be noted that different Keplerian elements have errors with very different orders of magnitude, resulting from the different magnitudes of the variations in the elements themselves. Finally, observe that the errors do not return to zero at the end of each revolution, due to the discrepancy between the frequency of the perturbed system and the frequency of the analytical solution. Correcting this discrepancy, by controlling the frequency of the power expansion (shown in the following section), will produce a solution with a similar short-term behavior of the error, but a better long-term one.

### C. Perturbation Method With Control in Frequency

*1. Construction of the Linear Operator*

Controlling the frequency of the expansion allows eliminating the secular terms appearing in the analytical solution, thus improving the long-term behavior of the solution generated by the operator matrix. The frequency is controlled through a Lindstedt-Poincaré expansion, that is, by rewriting the zeroth, first, and second order expansions of the equations of motion [Eqs. (38), (40), (41)] as a function of a time variable $\tau = \omega t$ with a frequency $\omega$ of the form

$$\omega = \omega_0 + \omega_1 J_2 + \omega_2 J_2^2 \qquad (45)$$

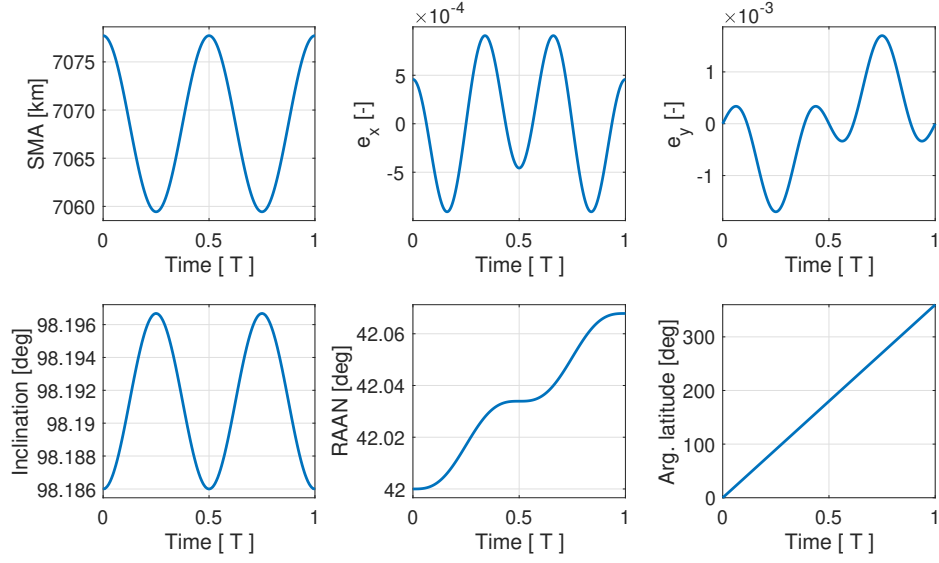

Fig. 2    Orbital elements evolution for a low-eccentricity frozen orbit

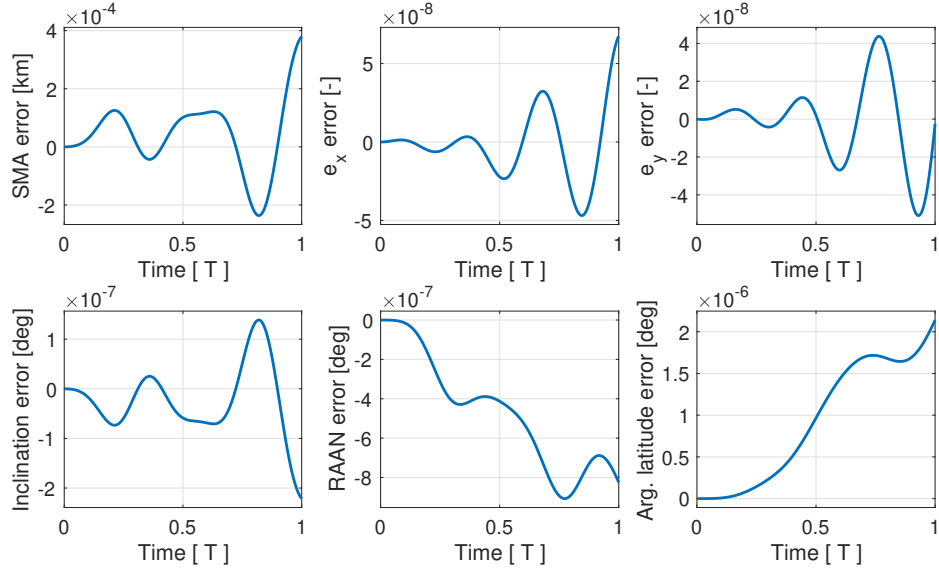

Fig. 3    Orbital elements error for a low-eccentricity frozen orbit, using an expansion without controlled frequency

The frequencies $\omega_0$, $\omega_1$, and $\omega_2$ are constants that depend on the initial conditions, therefore the time $t$ is a linear function of $\tau$. For example, in the case of $\beta$, this results in equations of motion with the form

$$\frac{d\beta_0}{d\tau} = \frac{1}{\omega_0} \left.\frac{d\beta}{dt}\right|_0$$

$$\frac{d\beta_1}{d\tau} = \frac{1}{\omega_0}\left(\left.\frac{d\beta}{dt}\right|_1 - \omega_1\frac{d\beta_0}{d\tau}\right)$$

$$\frac{d\beta_2}{d\tau} = \frac{1}{\omega_0}\left(\left.\frac{d\beta}{dt}\right|_2 - \omega_2\frac{d\beta_0}{d\tau} - \omega_1\frac{d\beta_1}{d\tau}\right) \tag{46}$$

where a vertical bar with a number, i.e. $(d\beta/dt)|_i$, denotes the $i^{th}$ order terms of $d\beta/dt$. The equations for the other elements are obtained by the same process. Analytically solving these equations of motion and selecting the frequencies to eliminate the secular terms in $\beta$, $X$, $Y$, and $\theta$ leads to

$$\omega_0 = \frac{\sqrt{\mu}}{R^{3/2}}\beta(t_0)^3$$

$$\omega_1 = \frac{3\sqrt{\mu}}{4R^{3/2}}\beta(t_0)^7\left(C_\theta(t_0)^2\left(3 - 3\beta(t_0)^2 p(t_0)^2\right) + \beta(t_0)^2 p(t_0)^2\left(3S_\theta(t_0)^2 + 8\right) - 3S_\theta(t_0)^2 - 2\right)$$

$$\omega_2 = \frac{3\sqrt{\mu}}{32R^{3/2}}\beta(t_0)^{11}\left(-6C_\theta(t_0)^2\left(\beta(t_0)^2 p(t_0)^2 - 1\right)\left(3\beta(t_0)^2 p(t_0)^2\left(13S_\theta(t_0)^2 + 25\right) - 39S_\theta(t_0)^2\right.\right.$$
$$\left. - 17\right) + 39C_\theta(t_0)^4\left(\beta(t_0)^2 p(t_0)^2 - 1\right)^2 - 2\beta(t_0)^2 p(t_0)^2\left(39S_\theta(t_0)^4 + 276S_\theta(t_0)^2 + 98\right)$$
$$\left. + \beta(t_0)^4 p(t_0)^4\left(39S_\theta(t_0)^4 + 450S_\theta(t_0)^2 + 325\right) + 39S_\theta(t_0)^4 + 102S_\theta(t_0)^2 + 51\right) \tag{47}$$

Now having the equations of motion [Eq. (46)] with frequencies selected to cancel the secular terms, it is finally possible to generate the operator matrix. To obtain an operator matrix independent of the initial conditions, besides the previously defined $k_\beta$, it is also necessary do define the variables $k_{\omega_1} = \omega_1/\omega_0$ and $k_{\omega_2} = \omega_2/\omega_0$, with time derivatives

$$\frac{dk_{\omega_1}}{d\tau} = \frac{dk_{\omega_2}}{d\tau} = 0 \tag{48}$$

Applying Algorithm 1 produces a linear operator matrix with size $625 \times 625$, describing the system as $d\mathbf{v}/d\tau = M\mathbf{v}$.

To analyze the generated operator, we first look at its eigenstructure (Fig. 4). The eigenvalues, computed using the Advanpix Multiprecision Computing Toolbox [45], are pure imaginary numbers, with only small real parts resulting from numerical errors in the computation of the eigenvalues. These eigenvalues result from the combination of the eigenvalues of the unperturbed system ($\pm i$ and 0) [8], therefore all have integer imaginary parts. The secular terms of the solution, namely in the evolution of $\Omega_1$ and $\Omega_2$, are associated with the defective eigenvalues (eigenvalues with algebraic multiplicity larger than the geometric multiplicity). The fact that all the eigenvalues are imaginary shows the numerical stability of the matrix, as it guarantees that the matrix will not lead initial state errors and numerical errors to grow exponentially, the latter being especially important when dealing with large matrices and long propagation times.

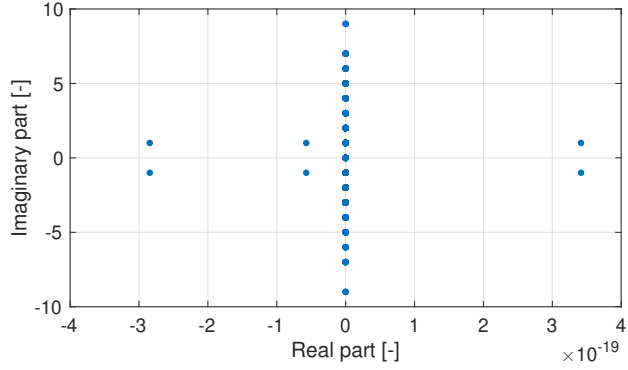

**Fig. 4 Eigenvalues of the operator matrix resulting from the expansion with controlled frequency**

*2. Frozen Orbit Conditions*

While analytically solving the equations of motion, we also find the conditions on the initial osculating elements that ensure a low-eccentricity frozen orbit. In Keplerian orbital elements these are

$$e_x(t_0) = \frac{J_2 R^2}{4a(t_0)^2} \cos(\theta(t_0)) \left(7 \cos(2\theta(t_0)) \sin^2(i(t_0)) + 11 \cos^2(i(t_0)) - 5\right)$$
$$e_y(t_0) = \frac{J_2 R^2}{4a(t_0)^2} \sin(\theta(t_0)) \left(7 \cos(2\theta(t_0)) + 14 \sin^2(\theta(t_0)) \cos^2(i(t_0)) - 1\right) \quad (49)$$

where the initial $x$ and $y$-eccentricities are given as functions of the initial semi-major axis, inclination, and argument of latitude. These conditions are different from the ones determined by Arnas [38], since different sets of elements were used, however the numerical results of the two coincide up to second order in $J_2$, as would be expected given that both are determined based on second-order approximations.

*3. Application*

To test the accuracy of this solution, we again consider the frozen sun-synchronous orbit with initial osculating elements $a = 7077.722$ km, $e_x = 4.5742 \times 10^{-4}$, $e_y = 0$, $i = 98.186$ deg, $\Omega = 42.0$ deg, and $\theta = 0$ deg. The evolution of the error in the analytical solution compared to the numerical integration is plotted over one revolution (1.645 hour) in Fig. 5. Observe how the errors in semi-major axis, eccentricities, and inclination return to approximately zero at the end of the revolution, leading the analytical solution to be extremely accurate even over long-term propagations (Fig. 6); after 100 revolutions, the maximum semi-major axis error has an order of magnitude of 0.1 m.

The long-term accuracy of the linear operator is also tested for a collection of frozen and non-frozen orbits. Using initial osculating elements $a = 7077.722$ km, $\Omega = 42.0$ deg, and $\theta = 0$ deg, the maximum position error observed over twenty revolutions is analyzed for orbits with varying initial inclinations and eccentricities (Fig. 7). Each curve in this plot represents multiple orbits; for instance, the orbit analyzed in Fig. 5 and Fig. 6 corresponds to the point in the blue curve with $i = 98.186$ deg. A similar analysis is done by instead fixing the initial osculating elements $i = 98.186$ deg, $\Omega$ = 42.0 deg, and $\theta$ = 0 deg, and plotting the maximum position error observed over twenty revolutions for orbits with varying initial semi-major axes and eccentricities (Fig. 8).

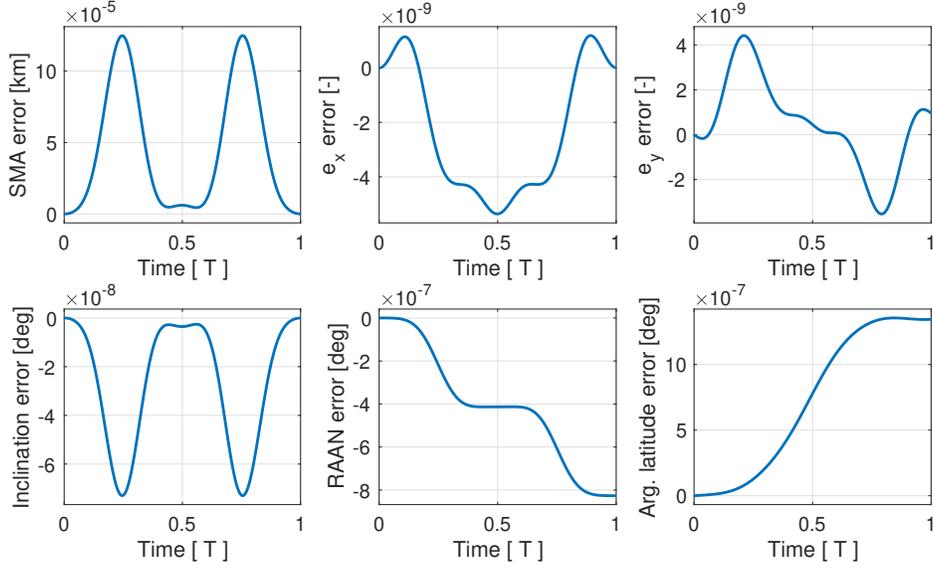

**Fig. 5** Orbital elements error for a low-eccentricity frozen orbit, using an expansion with controlled frequency

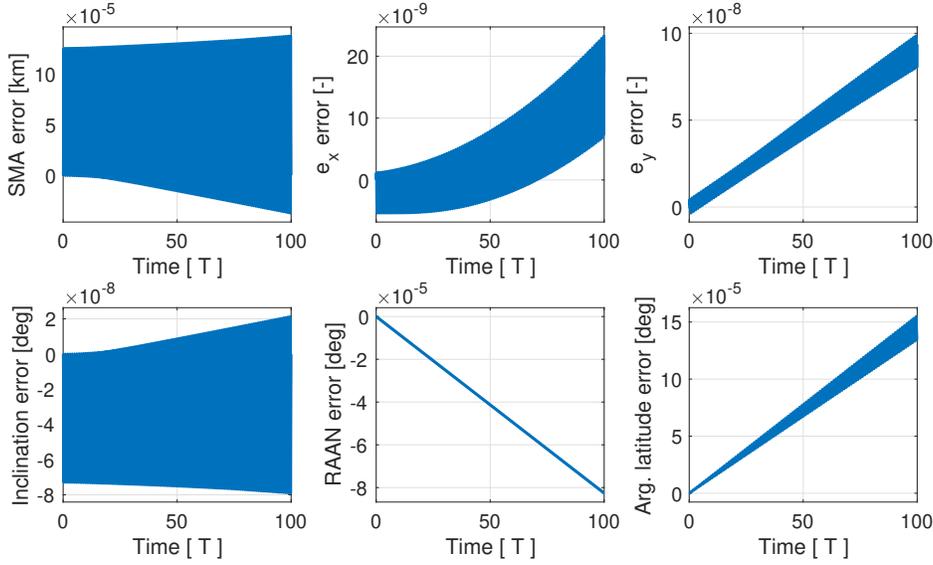

**Fig. 6** Orbital elements error for a low-eccentricity frozen orbit, using an expansion with controlled frequency

After one revolution both frozen and non-frozen orbits are approximated with a similar accuracy, corresponding to the order of the used perturbation method. However, when analyzing longer propagations, twenty revolutions in the shown examples, the error of the method is much smaller for frozen and near-frozen orbits. Since the solution is developed using a single frequency, which is characteristic of the dynamics in frozen orbits, the improvement in the long-term error behavior produced by the frequency control is largest for frozen and near-frozen orbits, with the error tending to grow as orbits deviate from the frozen condition. This is consistent with the expected behavior of the proposed methodology.

It is possible to observe large variations in the position accuracy of the solution when changing the initial osculating orbital elements, particularly visible when changing the initial inclination (Fig. 7), and present even when comparing frozen orbits. This accuracy variation is simply a result of differences in the magnitude of the terms neglected when executing the power expansion. In this case, the position accuracy is primarily limited by the accuracy of the argument of latitude. Consequently, variations in accuracy are mainly due to the magnitude of the terms neglected in the differential equation for this orbital element. Since the argument of latitude is the main source of position error, in some cases it may be beneficial to construct a "hybrid" solution, where this element is approximated with a higher-order expansion than the remaining orbital elements, thus improving the long-term accuracy of the approximated solution without unnecessarily increasing the size of the linear operator.

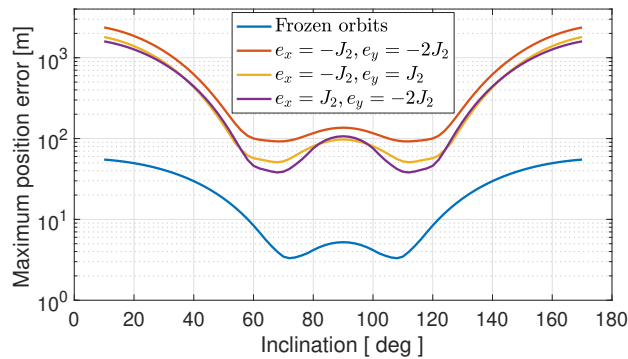

**Fig. 7** **Maximum position error over twenty revolutions, as a function of the initial inclination**

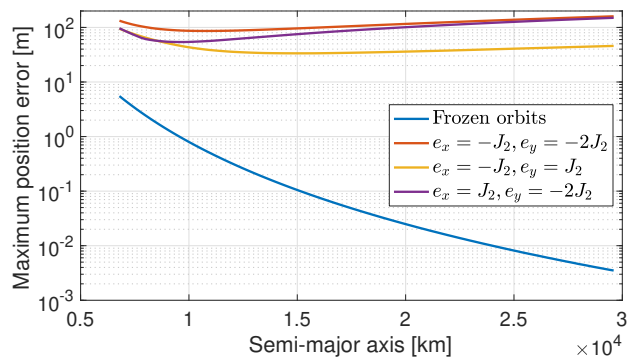

**Fig. 8** **Maximum position error over twenty revolutions, as a function of the initial semi-major axis**

### D. Comparison with the Koopman Operator

It is interesting to compare the performance of the proposed method with other options for obtaining a linear operator matrix. In particular, we analyze the Koopman operator [8, 9]. The order of the basis functions that the Koopman operator requires to represent a polynomial system of equations depends directly on the maximum exponent of those equations. This means that polynomials with larger exponents require more basis functions for the same accuracy. As a result, the direct application of the Koopman methodology to the equations of motion used here [Eq. (36)] becomes computationally intractable, due to the large exponents. As an alternative, the Koopman operator is instead generated using the equations of motion presented by Arnas and Linares [8], which have lower exponents. Since these equations are written as a function of an independent variable akin to the argument of latitude, we need to relate it to the time evolution used in this work, which is done using a transformation based on numerical integration.

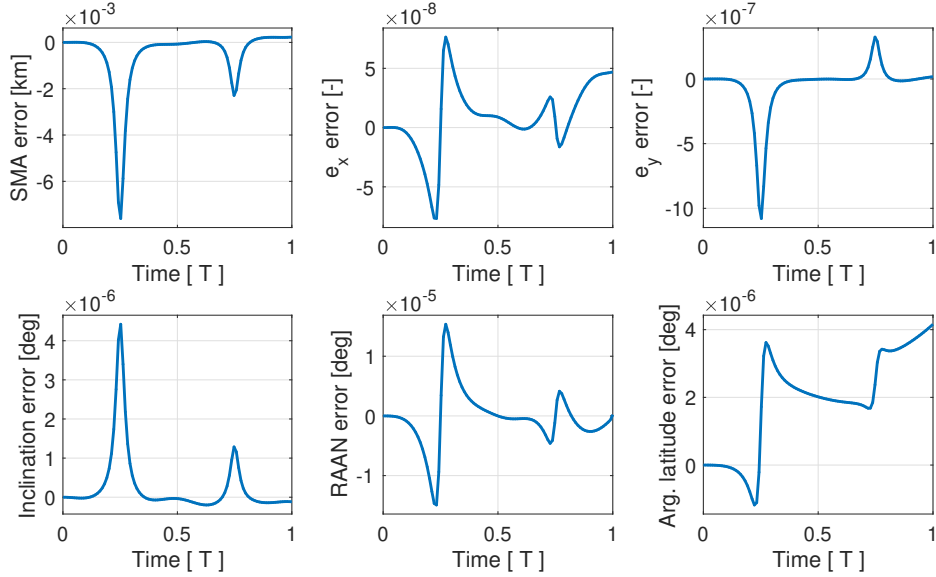

**Fig. 9   Orbital elements error for a low-eccentricity frozen orbit, using the Koopman operator**

To study the low-eccentricity frozen sun-synchronous orbit of the previous section, we generate the Koopman matrix using basis functions of order eleven (this order is not directly related to the order of the expansion used in this work, but rather to the order of the polynomials used to represent the solution). Compared with the proposed method based on the Lindstedt-Poincaré expansion, the Koopman operator requires a significantly larger matrix that is much slower to generate. Specifically, the proposed method requires a matrix of size $625 \times 625$ that can be computed in under a second in a common desktop. Conversely, the matrix produced by the Koopman operator has a size of $31825 \times 31825$ and requires more than three days to compute with the same hardware, due to the large amount of integrals required. Even though the Koopman matrix is much larger, the error (relative to the numerical integration) it generates over one revolution, plotted in Fig. 9, is one to two orders of magnitude larger than that from the second-order Lindstedt-Poincaré expansion (Fig. 5). Over a long-term propagation of 100 revolutions, the position error of the latter is again two orders of magnitude smaller (Fig. 10). The significantly smaller size of the operator generated in this work is related to the fact that it uses exactly the monomials necessary to represent a given order of the power expansion, while the Koopman operator also represents basis functions that do not contribute to the solution. In turn, this means that the method presented here can be applied to longer systems of equations before the size of the operator matrix and the computational time to obtain it start becoming problematic. A final advantage of the current method is that since it is based on a power expansion, changing the order has a very predictable effect on the error, which is not the case for the Koopman operator.

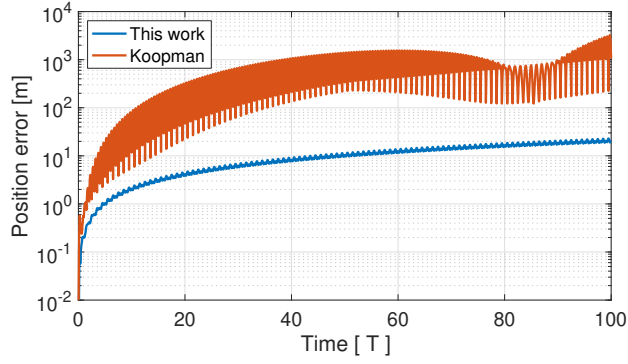

**Fig. 10  Position error for a low-eccentricity frozen orbit, using the operator with frequency control and the Koopman operator**

## V. Orbit Under $J_2$ Perturbation and Drag

As a final example, we apply the proposed method to approximate the dynamics under the effects of $J_2$ and atmospheric drag, showcasing the performance of the method when applied to non-conservative systems. Furthermore, a variation of the Lindstedt-Poincaré method with varying frequency is proposed, allowing the generated operator to dynamically adapt to the orbit decay induced by the drag.

For this example, the drag acceleration is modelled by

$$a_{drag} = -\frac{1}{2}\rho C_d \frac{S}{m} \|V\|^2 \frac{V}{\|V\|} \tag{50}$$

where $\rho$ is the atmospheric density, $S$ the cross-sectional area of the satellite, $m$ its mass, $C_d$ its drag coefficient, and $V$ the inertial velocity vector (static atmosphere assumption). Based on Eq. (35), the differential equation describing the evolution of each orbital element $x$ under the $J_2$ and drag perturbations is

$$\frac{dx}{dt} = \left.\frac{dx}{dt}\right|_{gravity} + \left.\frac{dx}{dt}\right|_{drag} \tag{51}$$

where the first term, representing the derivatives due to the point-mass gravity and $J_2$ term, is given by Eq. (36), and the second term is given by

$$\left.\frac{d\beta}{dt}\right|_{drag} = \frac{1}{2}\frac{\mu^{1/2}}{R^{1/2}}\rho C_d \frac{S}{m}\beta^2 \left(2\gamma - 1 + J_2^2\left(X^2 + Y^2\right)\right)^{1/2}$$

$$\begin{aligned}
\left.\frac{dX}{dt}\right|_{drag} &= -\frac{1}{J_2}\frac{\mu^{1/2}}{R^{1/2}}\rho C_d \frac{S}{m}\beta\left(C_\theta + J_2 X\right)\left(2\gamma - 1 + J_2^2\left(X^2 + Y^2\right)\right)^{1/2} \\
\left.\frac{dY}{dt}\right|_{drag} &= -\frac{1}{J_2}\frac{\mu^{1/2}}{R^{1/2}}\rho C_d \frac{S}{m}\beta\left(S_\theta + J_2 Y\right)\left(2\gamma - 1 + J_2^2\left(X^2 + Y^2\right)\right)^{1/2} \\
\left.\frac{dp}{dt}\right|_{drag} &= -\frac{1}{2}\frac{\mu^{1/2}}{R^{1/2}}\rho C_d \frac{S}{m}\beta p\left(2\gamma - 1 + J_2^2\left(X^2 + Y^2\right)\right)^{1/2} \\
\left.\frac{d\Omega}{dt}\right|_{drag} &= 0 \\
\left.\frac{dC_\theta}{dt}\right|_{drag} &= 0 \\
\left.\frac{dS_\theta}{dt}\right|_{drag} &= 0
\end{aligned} \quad (52)$$

Although these equations are not polynomial, they become polynomial when doing a series expansion with $J_2$ as the small parameter (using the binomial series), which allows applying the method for generating a linear operator.

## A. Perturbation Method With Constant Frequency

### 1. Construction of the Linear Operator

Similarly to Sec. IV.C, we obtain a matrix representing the $J_2$ problem with drag based on the application of the Lindstedt-Poincaré method. This expansion is done with two assumptions: that the density is constant, and that the drag perturbation is in the order of magnitude of $J_2^2$. The first assumption could be relaxed by considering the density to be polynomial with the altitude $H$, which is given by

$$H = \frac{R}{\beta^2\left(1 - J_2^2\left(X^2 + Y^2\right)\right)} - R \quad (53)$$

under the assumption of a spherical Earth. This polynomial density model could be defined, in particular, to correspond to the power expansion of the exponential density model. Even though introducing such a density model would allow a more realistic solution than the constant-density assumption, it would not alter the methodology to obtain the linear operator, simply generating one with more terms.

The assumption on the norm of the drag perturbation is specifically related to the magnitude of the term $\mu^{1/2}/R^{1/2}\rho C_d S/m$ in Eq. (52). Here we assume that drag is in the order of magnitude of $J_2^2$, but the same methodology could be applied if it was instead assumed to be in the order of $J_2$, which would be the case for very low altitude orbits, with the only difference being that different terms would appear in the solution. The assumption that the drag is in the order of $J_2^2$ is applied by multiplying each differential equation in Eq. (52) by $J_2^2 I_{J_2}^2$, where $I_{J_2}$ is a normalizing constant, defined as

$$I_{J_2} := \frac{1}{J_2} \quad (54)$$

Introducing this normalization eases the process of collecting the terms based on the powers of $J_2$ when doing the power expansion of each variable.

Similar to the previous sections, each orbital element is expanded as a power series with small parameter $J_2$

$$\beta \approx \beta_0 + \beta_1 J_2 + \beta_2 J_2^2$$
$$X \approx X_1 + X_2 J_2$$
$$Y \approx Y_1 + Y_2 J_2$$
$$p \approx p_0 + p_2 J_2^2$$
$$\Omega \approx \Omega_0 + \Omega_1 J_2 + \Omega_2 J_2^2$$
$$C_\theta \approx C_{\theta,0} + C_{\theta,1} J_2 + C_{\theta,2} J_2^2$$
$$S_\theta \approx S_{\theta,0} + S_{\theta,1} J_2 + S_{\theta,2} J_2^2 \tag{55}$$

In this case $p$ is also expanded, as in the presence of drag it is no longer a constant of motion. Note that its expansion only includes the zeroth ($p_0$) and second ($p_2$) order terms. Since we consider the drag to be in the order of $J_2^2$, the first order term has solution $p_1(t) = 0$, therefore including it in the expansion would increase the number of monomials, and consequently the size of the operator matrix, without altering the solution.

The frequency of the solution is again controlled through the Lindstedt-Poincaré expansion, using a regularization $\tau = \omega t$, with frequency $\omega = \omega_0 + \omega_1 J_2 + \omega_2 J_2^2$ given by Eq. (47), thus guaranteeing that $t$ is linear with $\tau$. The same frequency determined in the $J_2$ case is still sufficient to cancel all the secular terms appearing due to the application of the perturbation method, i.e. the secular terms representing the natural evolution of $\Omega$ (due to $J_2$), and of $\beta$ and $p$ (due to drag) are not canceled.

The expanded equations of motion are obtained via Eq. (46). Besides the variables $k_\beta$, $k_{\omega_1}$, and $k_{\omega_2}$, it is also useful to define $k_\rho = R/J_2^2 \rho C_d S/m$ with derivative $dk_\rho/d\tau = 0$, which allows obtaining an operator that does not depend on the characteristics of the spacecraft or the atmospheric density. Applying Algorithm 1 produces a matrix with size $635 \times 635$ describing the system linearly. Therefore, representing the effect of drag only requires a small increase in the dimensions of the associated matrix (representing just the effect of $J_2$ requires a $625 \times 625$ matrix, see Sec. IV.C).

*2. Application*

To study the accuracy of the generated operator, we consider the frozen sun-synchronous orbit from the previous sections, and a satellite with mass $m = 1285$ kg, cross-sectional area $S = 8.5$ m$^2$, drag coefficient $C_d = 2.2$, and constant density $\rho = 2 \times 10^{-14}$ kg m$^{-3}$. Over one revolution (1.645 hour), the operator generates a solution identical to Fig. 5. The solution is very accurate over short propagations but starts degrading when considering longer propagations, twenty revolutions in this case (Fig. 11). This quick error growth results from the discrepancy between the frequency of the true solution and of the analytical approximation. This occurs due to the effect of drag, which leads the semi-major axis to decay, with consequent increase in the frequency of the solution over time; meanwhile, the approximate solution was constructed using a constant frequency selected based on the initial state. It is worth highlighting that even though this solution is only accurate over short periods of time, it does allow obtaining the evolution of the orbital elements as a function of time in the presence of drag. To overcome the loss of accuracy for longer propagations, a method which allows the frequency of the operator to vary is proposed in the next section, however it will not provide a linear relationship between the independent variable and the time evolution.

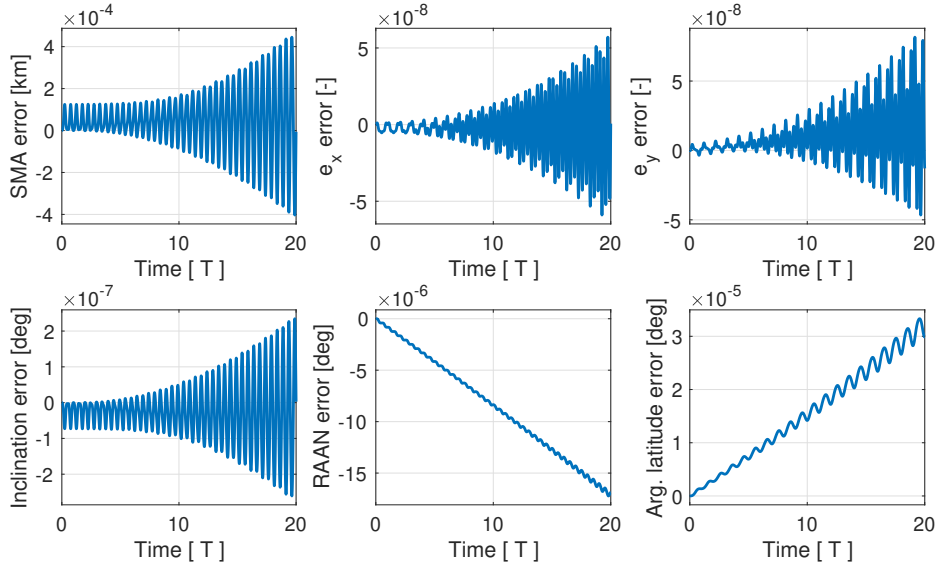

**Fig. 11    Orbital elements error for a low-eccentricity frozen orbit under the influence of drag and $J_2$**

### B. Perturbation Method With Varying Frequency

*1. Construction of the Linear Operator*

Approximating the orbit of a satellite subject to the $J_2$ and drag perturbations through the Lindstedt-Poincaré method, which is based on the definition of a constant frequency for the solution, was shown to provide limited long-term accuracy due to the variation of the frequency of the dynamics over time. To improve the long-term behavior of the solution, we propose a modification of the Lindstedt-Poincaré method that selects a varying frequency.

Similar to the Lindstedt-Poincaré method, the zeroth, first, and second order expansions of the equations of motion are written as a function of the independent variable $\tau$ through a regularization

$$\frac{d\tau}{dt} = \omega \tag{56}$$

However, contrary to the Lindstedt-Poincaré method, the frequency $\omega$, expanded as $\omega = \omega_0 + \omega_1 J_2 + \omega_2 J_2^2$, is selected to be a function of the orbital elements, therefore the relationship between $\tau$ and $t$ is not linear. Since the equation for the constant frequency determined using Lindstedt-Poincaré method (Eq. (47)) is valid for any initial condition, this equation is modified to produce an instantaneous frequency, computed based on the osculating orbital elements

$$\begin{aligned}\omega &= \frac{\sqrt{\mu}}{R^{3/2}}\beta^3 + J_2 \frac{3\sqrt{\mu}}{4R^{3/2}}\beta^7 \left(C_\theta^2 \left(3 - 3\beta^2 p^2\right) + \beta^2 p^2 \left(3S_\theta^2 + 8\right) - 3S_\theta^2 - 2\right) \\ &+ J_2^2 \frac{3\sqrt{\mu}}{32 R^{3/2}}\beta^{11} \left(-6C_\theta^2 \left(\beta^2 p^2 - 1\right)\left(3\beta^2 p^2 \left(13 S_\theta^2 + 25\right) - 39 S_\theta^2 - 17\right) + 39 C_\theta^4 \left(\beta^2 p^2 - 1\right)^2 \right.\\ &\left. - 2\beta^2 p^2 \left(39 S_\theta^4 + 276 S_\theta^2 + 98\right) + \beta^4 p^4 \left(39 S_\theta^4 + 450 S_\theta^2 + 325\right) + 39 S_\theta^4 + 102 S_\theta^2 + 51\right)\end{aligned} \quad (57)$$

Introducing the power expansion of the orbital elements [Eq. (55)] into this equation allows determining the zeroth, first, and second order terms of the frequency

$$\begin{aligned}\omega_0 &= \frac{\sqrt{\mu}}{R^{3/2}}\beta_0^3 \\ \omega_1 &= \frac{3\sqrt{\mu}}{4R^{3/2}}\beta_0^2 \left(\beta_0^7 p_0^2 \left(-3C_{\theta,0}^2 + 3 S_{\theta,0}^2 + 8\right) + \beta_0^5 \left(3 C_{\theta,0}^2 - 3 S_{\theta,0}^2 - 2\right) + 4\beta_1\right) \\ \omega_2 &= \frac{3\sqrt{\mu}}{32 R^{3/2}}\beta_0 \left(\beta_0^{14} p_0^4 \left(-18 C_{\theta,0}^2 \left(13 S_{\theta,0}^2 + 25\right) + 39 C_{\theta,0}^4 + 39 S_{\theta,0}^4 + 450 S_{\theta,0}^2 + 325\right) \right.\\ &\left. - 2\beta_0^{12} p_0^2 \left(-6 C_{\theta,0}^2 \left(39 S_{\theta,0}^2 + 46\right) + 39 C_{\theta,0}^4 + 39 S_{\theta,0}^4 + 276 S_{\theta,0}^2 + 98\right) - 48\beta_0^8 p_0^2 \left(C_{\theta,0} C_{\theta,1} - S_{\theta,0} S_{\theta,1}\right) \right.\\ &\left. - 72 \beta_1 \beta_0^7 p_0^2 \left(3 C_{\theta,0}^2 - 3 S_{\theta,0}^2 - 8\right) + 3\beta_0^{10} \left(-2 C_{\theta,0}^2 \left(39 S_{\theta,0}^2 + 17\right) + 13 C_{\theta,0}^4 + 13 S_{\theta,0}^4 + 34 S_{\theta,0}^2 + 17\right) \right.\\ &\left. + 48\beta_0^6 \left(C_{\theta,0} C_{\theta,1} - S_{\theta,0} S_{\theta,1}\right) + 56 \beta_1 \beta_0^5 \left(3 C_{\theta,0}^2 - 3 S_{\theta,0}^2 - 2\right) + 32 \beta_2 \beta_0 + 32 \beta_1^2\right)\end{aligned} \quad (58)$$

Having determined the frequency, the equations for the evolution of the orbital elements are obtained through Eq. (46). Note that in this case the equations for the frequencies are directly inserted into the equations of motion, instead of defining the frequencies as dependent variables when constructing the linear operator (as was done in the previous sections).

Unfortunately, the evolution of the time $t$ cannot be obtained directly from $\tau$, since the relationship between the two is not linear. Instead it is given by

$$\frac{dt}{d\tau} = \frac{1}{\omega} \quad (59)$$

Therefore the time evolution is now a dependent variable which can be approximated by a power expansion

$$t \approx t_0 + t_1 J_2 + t_2 J_2^2 \quad (60)$$

Introducing the power expansions of the orbital elements and of the time into the equation for $dt/d\tau$, and collecting the terms based on the order of $J_2$ allows obtaining the equations for $dt_0/d\tau$, $dt_1/d\tau$, and $dt_2/d\tau$, which can be solved with initial conditions $t_0(t_0) := t_0$ and $t_1(t_0) = t_2(t_0) = 0$.

Finally, and after defining $k_\beta$, $k_\rho$, and $k_{\omega_0} = 1/\omega_0$ (with $dk_{\omega_0}/d\tau = 0$) as auxiliary basis functions, applying Algorithm 1 produces a matrix with size $621 \times 621$, describing the system as $d\mathbf{v}/d\tau = M\mathbf{v}$, with both the orbital elements and the time included in the basis functions $\mathbf{v}$. This matrix $M$ does not depend on the properties of the spacecraft, atmospheric density, or main celestial body. Note that since the operator is constructed based on an osculating frequency, it is able to automatically adapt to changes in the frequency of the dynamics.

*2. Application*

This operator is tested using the same frozen sun-synchronous orbit and satellite from the previous sections. The evolution of the orbital elements error and time are plotted over 100 revolutions (164.5 hour), respectively in Fig. 12 and Fig. 13. The solution for the orbital elements is extremely accurate, with the error having the same order of magnitude obtained when considering just the $J_2$ perturbation (Fig. 6); for example, the maximum semi-major axis error is smaller than 0.25 m for the considered propagation time. Thus, introducing a varying frequency in the operator allows accurately approximating the changing frequency of the dynamics due to a dissipative perturbation.

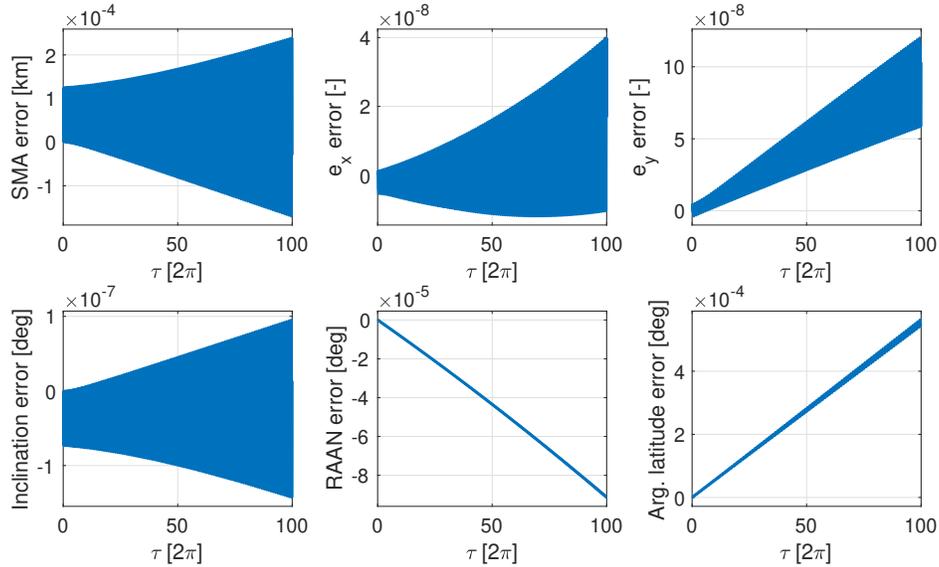

**Fig. 12** Orbital elements error for a low-eccentricity frozen orbit under drag and $J_2$, using an expansion with varying frequency

*3. Perturbation Method Based on the Argument of Latitude*

In the same way that in this work linear operators are generated based on the equations of motion with time as the independent variable, it is possible to instead use the argument of latitude as the independent variable. When obtaining an approximate solution to the $J_2$ problem based on a power expansion with the argument of latitude, there is no need to correct the frequency of the solution, as the dynamics evolve with the same frequency of the independent variable [38]. The same is true when generating a linear operator that also accounts for the drag perturbation. This operator (with size $572 \times 572$), tested using the same orbit and spacecraft from the previous sections propagated over 100 revolutions (Fig. 14), produces errors relative to the numerical solution with the same order of magnitude of Fig. 12 and Fig. 13, without the need to control the frequency of the solution.

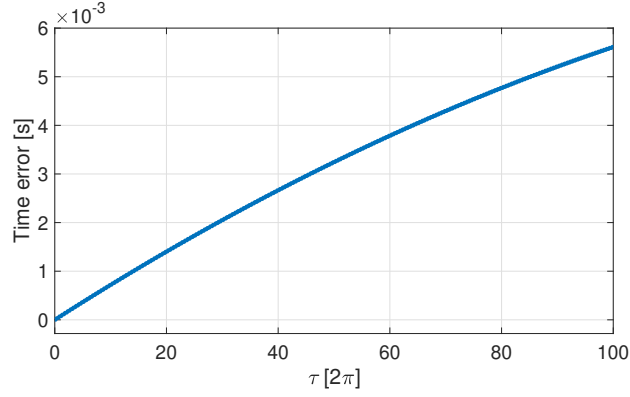

**Fig. 13  Time error for a low-eccentricity frozen orbit under drag and $J_2$, using an expansion with varying frequency**

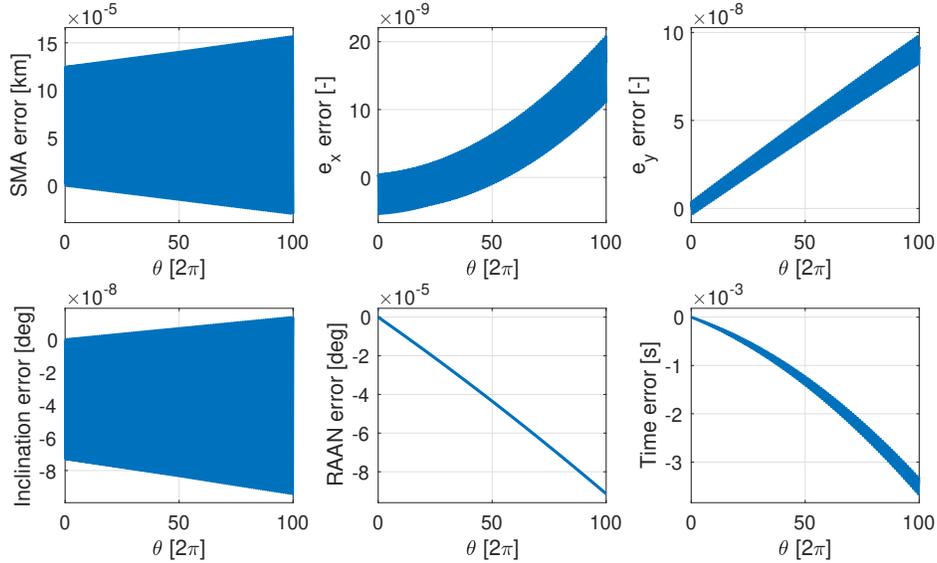

**Fig. 14  Orbital elements error for a low-eccentricity frozen orbit under drag and $J_2$, using an expansion without frequency control**

## VI. Conclusion

A new methodology for transforming nonlinear systems of perturbed differential equations into linear systems is introduced. This method is based on the application of a power-series perturbation expansion, producing an expanded system of polynomial differential equations, followed by the extension of the configuration space with the monomials that constitute those equations. The proposed methodology is based on osculating elements, meaning that the approximate solution directly provides the state of satellite as a function of time without the need of any additional transformations. Additionally, the resultant linear representation is completely independent of the initial conditions and therefore allows applying techniques developed for the study of linear systems, for example, in stability analysis, control, and estimation.

Since this operator is generated using classical perturbation theory, its accuracy can be easily adjusted by modifying the order used in the expansion. Furthermore, being based on the application of the Lindstedt-Poincaré expansion, the approximate solutions are stable over very long propagations times. Compared with the analytical Koopman operator in the $J_2$ problem, the proposed method generates a much smaller matrix ($625 \times 625$ instead of $31825 \times 31825$) in a much smaller time (tenths of seconds instead of days on the used testing hardware), which makes its use more practical and efficient. Moreover, even though our matrix is much smaller, when analyzing frozen orbits it is one to two orders of magnitude more accurate than the Koopman operator.

The proposed methodology is general and can be applied to a variety of problems, both conservative (in this case, the Duffing oscillator and the $J_2$ problem) and non-conservative (in this case, the $J_2$ problem with drag). When using a simple power expansion, this method can be applied to any perturbed system of differential equations that has a polynomial representation and that satisfies the presented conditions for a finite linear representation. Examples includes the zonal harmonics problem with higher order terms and the motion near the Lagrange points of the circular restricted three-body problem. To guarantee the long-term stability of the approximate solution, the method should be applied using a Lindstedt-Poincaré expansion, which additionally requires that the motion being described is characterized by a single frequency, thus implying the existence of oscillatory solutions (though not necessarily periodic). When applied to other systems, the resultant linear operators always have the accuracy characteristics of the used perturbation method, in this case, the Lindstedt-Poincaré expansion. Over one revolution the linear operator presents the accuracy associated with the selected perturbation order, over longer propagations the behavior of the error depends on the specific characteristics of the system being studied. In particular, the error growth can be expected to be slower for linearly stable systems than for unstable ones.

[4] Brunton, S. L., Brunton, B. W., Proctor, J. L., and Kutz, J. N., "Koopman Invariant Subspaces and Finite Linear Representations of Nonlinear Dynamical Systems for Control," *PLOS ONE*, Vol. 11, No. 2, 2016. https://doi.org/10.1371/journal.pone.0150171.

[5] Surana, A., "Koopman Framework for Nonlinear Estimation," *The Koopman Operator in Systems and Control*, edited by A. Mauroy, I. Mezić, and Y. Susuki, Springer Nature, Cham, 2020, pp. 59–79. https://doi.org/10.1007/978-3-030-35713-9_3.

[6] Servadio, S., Arnas, D., and Linares, R., "Dynamics Near the Three-Body Libration Points via Koopman Operator Theory," *Journal of Guidance, Control, and Dynamics*, Vol. 45, No. 10, 2022, pp. 1800–1814. https://doi.org/10.2514/1.G006519.

[7] Chen, T., and Shan, J., "Koopman-Operator-Based Attitude Dynamics and Control on SO(3)," *Journal of Guidance, Control, and Dynamics*, Vol. 43, No. 11, 2020, pp. 2112–2126. https://doi.org/10.2514/1.G005006.

[8] Arnas, D., and Linares, R., "Approximate Analytical Solution to the Zonal Harmonics Problem Using Koopman Operator Theory," *Journal of Guidance, Control, and Dynamics*, Vol. 44, No. 11, 2021, pp. 1909–1923. https://doi.org/10.2514/1.G005864.

[9] Arnas, D., "Solving Perturbed Dynamic Systems Using Schur Decomposition," *Journal of Guidance, Control, and Dynamics*, Vol. 45, No. 12, 2022, pp. 2211–2228. https://doi.org/10.2514/1.G006726.

[10] Servadio, S., Parker, W., and Linares, R., "Uncertainty Propagation and Filtering via the Koopman Operator in Astrodynamics," *Journal of Spacecraft and Rockets*, Vol. 60, No. 5, 2023, pp. 1639–1655. https://doi.org/10.2514/1.A35688.

[11] Servadio, S., Armellin, R., and Linares, R., "Koopman-Operator Control Optimization for Relative Motion in Space," *Journal of Guidance, Control, and Dynamics*, Vol. 46, No. 11, 2023, pp. 2121–2132. https://doi.org/10.2514/1.G007217.

[12] Brouwer, D., "Solution of the Problem of Artificial Satellite Theory Without Drag," *The Astronomical Journal*, Vol. 64, No. 1274, 1959, pp. 378–396. https://doi.org/10.1086/107958.

[13] Garfinkel, B., "The Orbit of a Satellite of an Oblate Planet," *The Astronomical Journal*, Vol. 64, No. 9, 1959, pp. 353–366. https://doi.org/10.1086/107956.

[14] Kozai, Y., "The Motion of a Close Earth Satellite," *The Astronomical Journal*, Vol. 64, No. 1274, 1959, pp. 367–377. https://doi.org/10.1086/107957.

[15] Kozai, Y., "Second-Order Solution of Artificial Satellite Theory without Air Drag," *The Astronomical Journal*, Vol. 67, No. 7, 1962, pp. 446–461. https://doi.org/10.1086/108753.

[16] Lyddane, R. H., "Small Eccentricities or Inclinations in the Brouwer Theory of the Artificial Satellite," *The Astronomical Journal*, Vol. 68, No. 8, 1963, pp. 555–558. https://doi.org/10.1086/109179.

[17] Cohen, C. J., and Lyddane, R. H., "Radius of Convergence of Lie Series for Some Elliptic Elements," *Celestial Mechanics*, Vol. 25, 1981, pp. 221–234. https://doi.org/10.1007/BF01228961.

[18] Lara, M., "Brouwer's Satellite Solution Redux," *Celestial Mechanics and Dynamical Astronomy*, Vol. 133, No. 47, 2021. https://doi.org/10.1007/s10569-021-10043-7.
37